\documentclass[11pt,a4paper]{article} 
\pdfoutput=1
\usepackage{jcappub}
\usepackage{enumerate}
\usepackage{epsfig} 
\usepackage{float} 
\usepackage[caption = false]{subfig}
\usepackage{wasysym} 
\usepackage{mathrsfs} 
\usepackage{amsfonts} 
\usepackage{amsbsy} 
\usepackage{amscd} 
\usepackage{pstricks} 
\usepackage{multirow} 
\usepackage{tikz}
\usepackage{color}
\usepackage{slashed}
\usepackage{array}
\usepackage{tablefootnote} 
\usepackage{enumitem}
\usepackage[utf8]{inputenc}
\usepackage{hyperref}
\usepackage{color}
\definecolor{myred}{rgb}{0.6,0,0} %usage:  {\textcolor{myred}{Hello World}}
\definecolor{myblue}{rgb}{0,0.2,0.4}
\definecolor{mygreen}{rgb}{0,0.9,0.1}
\definecolor{hc}{rgb}{.9,0.1,0.7}
\definecolor{hcout}{rgb}{.9,0.7,0.9}
\definecolor{Orange}{rgb}{0,0.2,0.9}
\makeatletter

\newcommand{\fmslash}[2][0mu]{%
  \mathchoice
    {\fmsl@sh\displaystyle{#1}{#2}}%
    {\fmsl@sh\textstyle{#1}{#2}}%
    {\fmsl@sh\scriptstyle{#1}{#2}}%
    {\fmsl@sh\scriptscriptstyle{#1}{#2}}}
\newcommand{\fmsl@sh}[3]{%
  \m@th\ooalign{$\hfil#1\mkern#2/\hfil$\crcr$#1#3$}}
\makeatother

\newcommand{\lsim}{{\;\raise0.3ex\hbox{$<$\kern-0.75em\raise-1.1ex\hbox{$\sim$}}\;}}
\newcommand{\gsim}{{\;\raise0.3ex\hbox{$>$\kern-0.75em\raise-1.1ex\hbox{$\sim$}}\;}}

\newcommand{\overbar}[1]{\mkern 1.5mu\overline{\mkern-1.5mu#1\mkern-1.5mu}\mkern 1.5mu}

\usepackage{array}
\newcolumntype{C}[1]{>{\centering\arraybackslash$}p{#1}<{$}}

\usepackage{bm} % bold in math mode
\newcommand{\be}{\begin{equation}}
\newcommand{\ee}{\end{equation}}
\newcommand{\bes}{\begin{equation*}}
\newcommand{\ees}{\end{equation*}}
\newcommand{\bea}{\begin{eqnarray}}
\newcommand{\eea}{\end{eqnarray}}
\newcommand{\beas}{\begin{eqnarray*}}
\newcommand{\eeas}{\end{eqnarray*}}
\title{MeV to multi-TeV thermal WIMPs: most conservative limits} 

\author[a]{Koushik Dutta,}
\author[b]{Avirup Ghosh,} 
\author[c,d]{Arpan Kar,} 
\author[a]{Biswarup Mukhopadhyaya}

\affiliation[a]{Department of Physical Sciences, Indian Institute of Science Education and Research, Kolkata, Mohanpur - 741246, India.}  
\affiliation[b]{School of Physical Sciences, Indian Association for the Cultivation of Science, 2A and 2B Raja S.C. Mullick Road, Kolkata 700 032.}
\affiliation[c]{Center for Quantum Spacetime, Sogang University, Seoul 121-742, South Korea.}
\affiliation[d]{Department of Physics, Sogang University, Seoul 121-742, South Korea.}

\emailAdd{koushik@iiserkol.ac.in}
\emailAdd{avirup.ghosh1993@gmail.com}
\emailAdd{arpankarphys@gmail.com} 
\emailAdd{biswarup@iiserkol.ac.in}

\abstract{We consider a weakly interacting massive 
particle (WIMP) dark matter (DM) annihilating into 
all possible Standard Model (SM) particle pairs, 
including the SM neutrinos, via $s$-wave processes and 
derive the branching ratio independent upper limit 
on the total annihilation cross-section 
$\langle \sigma v \rangle$ using the data of 
CMB, gamma-ray, cosmic-ray and several neutrino 
observations. For conservative choices of all relevant 
astrophysical parameters, we obtain upper limits 
of $10^{-23}-10^{-25}\,{\rm cm}^3{\rm s}^{-1}$ 
on the total $\langle \sigma v \rangle$ for the WIMP mass 
range $10\,{\rm MeV}-100\,{\rm TeV}$, thus making the 
entire mass range consistent with the observed relic density. 
An important input that goes into our analysis 
is the assumption that thermal WIMPs can 
have significant coupling to the SM neutrinos.} 

\subheader{{\normalfont{CQU}}{\Large{e}}{\normalfont{ST-2022-0706}}}
\date{\today}


\begin{document}
\maketitle
\newpage

%******************************************************* 
%\noindent
\section{Introduction} %\\
Weakly interacting massive particles (WIMPs) 
%have been widely studied as the dark matter (DM) 
%of our Universe. WIMPs 
are widely studied DM candidates of our 
Universe. A WIMP maintains thermal equilibrium 
with the Standard Model (SM) plasma in the 
Early Universe and freezes-out when its annihilation 
rate falls below the expansion rate of the 
Universe~\cite{Steigman:2012nb}. 
The relic density requirement, 
$\Omega_\chi h^2 = 0.12$~\cite{Steigman:2012nb,Planck:2018vyg}, 
specifies its thermally averaged total annihilation cross-section, 
$\langle \sigma v \rangle \sim 10^{-26}\,{\rm cm}^{3}\,{\rm s}^{-1}$, 
at the time of freeze-out, but does not fix the branching 
ratio (BR) of each individual annihilation channel. 
The observed structure of the Universe puts a lower 
bound of $\mathcal{O}({\rm keV})$ on the mass ($m_\chi$) 
of a thermal DM~\cite{Irsic:2017ixq}, while, the 
upper bound %on $m_\chi$ 
comes from the unitarity of the S-matrix which disallows 
$m_\chi \gtrsim \mathcal{O}(100\,{\rm TeV})$~\cite{Griest:1989wd,Smirnov:2019ngs}. 
Therefore, any value of $m_\chi$ lying in the range 
MeV - 100 TeV is \textit{prima facie} allowed unless 
it is ruled out by some existing observations. 

Although direct detection 
experiments~\cite{Aprile:2018dbl,Akerib:2016vxi,Cui:2017nnn,Agnes:2018ves,PhysRevD.100.102002,Amole:2019fdf,Schumann:2019eaa} %Abdelhameed:2019hmk
and collider 
searches~\cite{Sirunyan:2017hci,CMSlim,ATLASlim,Khachatryan:2016whc,Abdallah:2015ter} 
have put limits on thermal WIMPs, 
assumptions on their interactions enter into the 
derivations of such constraints (see, 
for example,~\cite{Arcadi:2017kky,GAMBIT:2021rlp}). 
Similarly, while analyzing potential WIMP signals in indirect searches, 
too, assumptions on their annihilation branching fractions 
are 
implicit~\cite{Planck:2015fie,Arcadi:2017kky,GAMBIT:2021rlp,Planck:2018vyg,Fermi-LAT:2016uux,Lu:2015pta,John:2021ugy,Profumo:2016idl,Arguelles:2019ouk,Egorov:2022rhb}. 
It is on the basis of such assumptions, a WIMP of 
mass up to \textit{a few hundreds of GeV} is claimed 
to be ruled out~\cite{Fermi-LAT:2016uux,Lu:2015pta,John:2021ugy}, 
since its maximum allowed $\langle \sigma v \rangle$ falls 
below $10^{-26}\,{\rm cm}^{3}\,{\rm s}^{-1}$, the value 
required from relic density.

%\sout{However, such channel specific constraints %are not 
%universally applicable. This is because, DM %particles, 
%may, in principle, annihilate into multiple %channels 
%with arbitrary BRs, in which case the resulting %upper 
%limit on total $\langle \sigma v \rangle$ can be 
%quite different.} 
Note that, such channel specific constraints are not 
universally applicable. This is because, 
in a given DM model, DM particles %may in principle 
can annihilate into multiple channels with BRs 
dictated by the model parameters, and in such cases the resulting 
upper limits on total $\langle \sigma v \rangle$ can be quite 
different. However, due to the existence of a large number of DM 
models it is impossible to perform a study on a model-by-model basis 
and hence, the most general approach to meaningfully interpret the 
impact of the observational data on WIMP DM annihilations is 
to vary the branching ratios of each channel arbitrarily. 
Following this approach, Ref.~\cite{Leane:2018kjk} claimed 
that thermal WIMPs annihilating into visible SM final states 
via $2 \rightarrow 2$ $s$-wave processes were disallowed 
for $m_\chi \lesssim\,20\,{\rm GeV}$ by the 
data available till then. On the other hand, 
while Big-Bang Nucleosynthesis (BBN) and 
Cosmic Microwave Background (CMB) data rule out 
$m_\chi \lesssim 10\,{\rm MeV}$~\cite{Nollett:2013pwa,Nollett:2014lwa} 
(Ref.~\cite{Kawasaki:2015yya} suggests 
stronger BBN constraints under specific assumptions),  
$m_\chi$ values much smaller than $20\,{\rm GeV}$ are in principle 
permitted if the DM $\chi$ annihilates substantially 
into SM 
neutrinos~\cite{Beacom:2006tt,Yuksel:2007ac,Arguelles:2019ouk}. 
%\sout{In this work, we have taken a general approach, without 
%any bias on DM annihilation branching ratios and concluded 
%that the entire $m_\chi$ range 10 MeV - 100 TeV 
%is generally allowed, after taking into account the most 
%updated constraints from the neutrino  
%observations~\cite{Super-Kamiokande:2020sgt,Super-Kamiokande:2021jaq,IceCube:2017rdn,ANTARES:2019svn,Albert:2016emp,Iovine:2019rmd,ANTARES:2020leh} as well as 
%the CMB~\cite{Planck:2015fie,Planck:2018vyg}, 
%$\gamma$-ray~\cite{Fermi-LAT:2016uux,HESS:2022ygk} and 
%cosmic-ray~\cite{AMS:2019rhg} data. The crucial role of the 
%BR of $\nu\bar{\nu}$ is especially notable here.} 

%\sout{In this work, we re-examine the above claim %in 
%the most general approach, by removing all bias 
%about WIMP DM annihilation branching ratios, 
%excepting the postulate that the dominant %annihilations 
%proceed via two-body SM final states.} 
In this work, we re-examine the above claim in 
the most general approach, by considering the possibility of 
WIMP DM annihilations into SM neutrinos, and only assuming that 
the dominant annihilations proceed via two-body SM final 
states. Our conclusion is that the entire $m_\chi$ 
range 10 MeV - 100 TeV can be allowed for such WIMP 
candidates. This conclusion is arrived at, after taking 
into account the most updated constraints from the relevant 
neutrino observations~\cite{Super-Kamiokande:2020sgt,Super-Kamiokande:2021jaq,IceCube:2017rdn,ANTARES:2019svn,Albert:2016emp,Iovine:2019rmd,ANTARES:2020leh} 
as well as the CMB~\cite{Planck:2015fie,Planck:2018vyg}, 
$\gamma$-ray~\cite{Fermi-LAT:2016uux,HESS:2022ygk} 
and cosmic-ray~\cite{AMS:2019rhg} data. 
Note that, while we are not tilting towards 
any particular DM model, it turns out that 
%the most conservative limits for 
WIMPs of $m_\chi \lesssim \mathcal{O}(10\,{\rm GeV})$ 
are allowed if they have substantial branching 
ratios into SM neutrinos. Although such a feature is 
not very common, it is phenomenologically viable. 
There exist scenarios where DM lighter than 
$\sim 10\,{\rm GeV}$ dominantly annihilates into SM 
neutrinos~\cite{Blennow:2019fhy,Ballett:2019pyw,Dutta:2019fxn,ElAisati:2017ppn,Farzan:2011ck} 
and such models also fall under the purview of the 
present study. 
On the whole, the present study has the following 
novel features:
\begin{itemize}
\item The possibility of DM annihilating into 
$\nu\bar{\nu}$ pairs has been included in an 
unbiased manner. This affects our analysis in 
three ways. First, it allows arbitrary reduction 
in the branching ratios into 
\textit{all SM particle pairs other than neutrinos}. 
Secondly, the updated constraints ensuing from all 
neutrino observation experiments~\cite{Super-Kamiokande:2020sgt,Super-Kamiokande:2021jaq,IceCube:2017rdn,ANTARES:2019svn} 
have been included, especially for cases 
where the neutrino-pair channel has 
appreciable branching ratio. 
And thirdly, for high-mass DM, the energetic 
neutrinos may have enhanced W-emission rates, 
resulting in additional contributions to 
high-energy $\gamma$-ray and cosmic-ray events, 
whose effects on the current constraints have 
been calculated. 
%for different values of the 
%$\nu\bar{\nu}$ branching ratios.
\item The most updated data from 
Planck~\cite{Planck:2018vyg}, 
Fermi-LAT~\cite{Fermi-LAT:2016uux} and 
AMS-02~\cite{AMS:2019rhg} have been used, 
once more without any {\it a priori} assumption regarding the 
values of the %theoretical bias regarding 
DM annihilation branching ratios.
\item The H.E.S.S. data~\cite{HESS:2022ygk} and 
their relevance for multi-TeV WIMP annihilations 
have been taken into account.
\end{itemize}

%In this work, we re-examine the above-mentioned 
%claim in the most general approach, by relaxing 
%all possible biases on DM annihilation branching 
%ratios and concluded that the entire $m_\chi$ range 
%10 MeV - 100 TeV is generally allowed, after taking 
%into account the most updated constraints from the 
%relevant neutrino observations~\cite{Super-Kamiokande:2020sgt,Super-Kamiokande:2021jaq,IceCube:2017rdn,ANTARES:2019svn,Albert:2016emp,Iovine:2019rmd,ANTARES:2020leh} as well as 
%the CMB~\cite{Planck:2015fie,Planck:2018vyg}, 
%$\gamma$-ray~\cite{Fermi-LAT:2016uux,HESS:2022ygk} and cosmic-ray~\cite{AMS:2019rhg} data. The crucial role of 
%the BR of $\nu\bar{\nu}$ is especially notable here.
%Therefore, our work goes beyond the earlier studies 
%in several aspects: 
%\begin{itemize}
%\item DM annihilations to $\nu\bar{\nu}$ (with arbitrary BRs) 
%is included.
%\item Most updated data of Planck~\cite{Planck:2018vyg}, Fermi-LAT~\cite{Fermi-LAT:2016uux} and AMS-02~\cite{AMS:2019rhg} have been 
%used.
%\item Relevance of H.E.S.S data~\cite{HESS:2022ygk} 
%in constraining multi-TeV WIMP annihilations are pointed out.
%\item Data of all relevant neutrino 
%observations~\cite{Super-Kamiokande:2020sgt,Super-Kamiokande:2021jaq,IceCube:2017rdn,ANTARES:2019svn} are taken 
%into account.
%\end{itemize}
%}

This paper is organized as follows: in Sec.~\ref{sec:sec2} 
we describe the limits obtained from CMB, $\gamma$-ray 
and cosmic-ray data. Sec.~\ref{sec:sec3} is 
devoted to the discussion of the limits coming from the 
neutrino observations. In Sec.~\ref{sec:sec4} we present the 
constraints we have obtained on the total annihilation 
cross-section using all observational data. Finally, we 
conclude in Sec.~\ref{sec:sec5}. 

%\vspace{1mm}
%\noindent
\section{Limits from CMB, $\gamma$-ray and cosmic-ray observations}
\label{sec:sec2}
%\\
% \sout{DM annihilations inside the astrophysical objects produce 
% SM particle pairs ${\rm SM}_1\overbar{\rm SM}_2$ 
% (i.e., $e^+e^-$, $\mu^+\mu^-$, $\tau^+\tau^-$, 
% $b\bar{b}$, $t\bar{t}$, $q\bar{q}$, $gg$, $W^+W^-$, 
% $ZZ$, $\gamma\gamma$, $hh$, $h\gamma$, $Z\gamma$, 
% $Zh$, $\nu\bar{\nu}$)
% \footnote{For $\nu\bar{\nu}$ final state, $\nu_e\bar{\nu}_e$, 
% $\nu_\mu\bar{\nu}_\mu$ and $\nu_\tau\bar{\nu}_\tau$, 
% while, for $q\bar{q}$ channel, $u\bar{u}$, $d\bar{d}$, 
% $c\bar{c}$ and $s\bar{s}$ are assumed to have equal BRs, 
% so that their sums represent the total branching 
% fractions attributed to $\nu\bar{\nu}$ and $q\bar{q}$, 
% respectively.} which give rise to photons, stable charged 
% (anti)matter particles and (anti)neutrinos via cascade decays. 
% For a particular choice of the relevant astrophysical 
% parameters, fluxes of such stable SM particles are 
% proportional to the total $\langle \sigma v \rangle$ 
% for any given $m_\chi$.} 
%
%
Fluxes of $\gamma$, $e^-(e^+)$, $\nu(\bar{\nu})$ 
produced from the cascade decays of the DM 
annihilation induced SM particle pairs, 
${\rm SM}_1\overbar{\rm SM}_2$ 
(i.e., $e^+e^-$, $\mu^+\mu^-$, $\tau^+\tau^-$, 
$b\bar{b}$, $t\bar{t}$, $q\bar{q}$, $gg$, $W^+W^-$, 
$ZZ$, $\gamma\gamma$, $hh$, $h\gamma$, $Z\gamma$, 
$Zh$ and $\nu\bar{\nu}$)
\footnote{For $\nu\bar{\nu}$ final state, $\nu_e\bar{\nu}_e$, 
$\nu_\mu\bar{\nu}_\mu$ and $\nu_\tau\bar{\nu}_\tau$, 
while, for $q\bar{q}$ channel, $u\bar{u}$, $d\bar{d}$, 
$c\bar{c}$ and $s\bar{s}$ are assumed to have equal BRs, 
so that their sums represent the total branching 
fractions attributed to $\nu\bar{\nu}$ and $q\bar{q}$, 
respectively.}, are proportional to the total 
$\langle \sigma v \rangle$, for a fixed $m_\chi$ 
and a given set of relevant astrophysical parameters. 
These fluxes are compared with the observed data 
to derive limits on the 
$\langle \sigma v \rangle - m_\chi$ plane. 
Here, such fluxes are obtained using 
\texttt{Pythia}~\cite{Sjostrand:2014zea,Christiansen:2014kba}, 
with the only exceptions being sub-GeV DM annihilations, 
for which we use the available analytic 
formulae~\cite{Coogan:2019qpu,Cirelli:2020bpc}.
%\red{\textbf{Event generator discussion and Majorana particle DM}}
Instead of \texttt{Pythia} one can use other monte-carlo event 
generators~\cite{Bellm:2015jjp,Bellm:2017bvx,Sherpa:2019gpd} 
to obtain the DM induced fluxes of stable SM particles. 
Such fluxes can be sensitive to the choice of the event 
generator~\cite{Cembranos:2013cfa,Cirelli:2010xx,Niblaeus:2019ldk}, 
and depending on the event generator used, the 
DM induced $e^-(e^+)$ and $\nu(\bar{\nu})$ fluxes may vary 
up to $\sim 20\%$, while the gamma-ray fluxes may change by 
a factor of a few. Such differences are attributed to the 
QED and electroweak corrections incorporated in 
\texttt{Pythia}. For annihilation channels leading to hadronic 
end products like antiprotons, the differences can be even larger 
and are caused by the differences in the hadronization modelling 
used by different generators. In this work, we have not 
considered such dependencies of the DM annihilation fluxes on 
the event generators and assumed that these fluxes are 
faithfully simulated using \texttt{Pythia}.

Throughout the paper, we shall consider a 
self-conjugate WIMP DM candidate (e.g., real 
scalar or Majorana fermion), for illustration. 
However, it is straightforward to extend our 
analysis to other possible WIMP scenarios. 
In this section, we discuss the constraints 
obtained from the non-observation of DM induced 
$\gamma$ and $e^-(e^+)$ fluxes. 
% SM particle pairs ${\rm SM}_1\overbar{\rm SM}_2$ 
% (i.e., $e^+e^-$, $\mu^+\mu^-$, $\tau^+\tau^-$, 
% $b\bar{b}$, $t\bar{t}$, $q\bar{q}$, $gg$, $W^+W^-$, 
% $ZZ$, $\gamma\gamma$, $hh$, $h\gamma$, $Z\gamma$, 
% $Zh$, $\nu\bar{\nu}$)
% \footnote{For $\nu\bar{\nu}$ final state, $\nu_e\bar{\nu}_e$, 
% $\nu_\mu\bar{\nu}_\mu$ and $\nu_\tau\bar{\nu}_\tau$, 
% while, for $q\bar{q}$ channel, $u\bar{u}$, $d\bar{d}$, 
% $c\bar{c}$ and $s\bar{s}$ are assumed to have equal BRs, 
% so that their sums represent the total branching 
% fractions attributed to $\nu\bar{\nu}$ and $q\bar{q}$, 
% respectively.}, produced from DM annihilations 
% inside the astrophysical objects give rise to 
% $\gamma$, $e^-(e^+)$, $\nu(\bar{\nu})$ via 
% cascade decays. For a given set of relevant 
% astrophysical parameters and a fixed $m_\chi$, 
% fluxes of such stable SM particles are 
% proportional to the total $\langle \sigma v \rangle$. 

%\sout{Constraints from the observations, which are sensitive 
%to $\gamma$ and $e^-(e^+)$, are discussed below:}

%Constraints obtained from the non-observation 
%of DM induced $\gamma$ and $e^-(e^+)$ fluxes are discussed 
%below.

\vspace{0.1cm}

\noindent
\subsection{Planck} 

Ionizing particles (mainly $e^-$, $e^+$ and $\gamma$) 
originating from DM annihilations, can change the ionization 
history of the hydrogen and helium gasses, thereby perturbing 
CMB anisotropies. 
%inject energies to CMB and perturb its spectra. 
%$\gamma$ and $e^-(e^+)$ 
%produced from DM annihilations, 
%inject energies to CMB and perturb its spectra. 
%Following~\cite{Slatyer:2015jla}, 
%We use the CMB anisotropy observed by 
Planck~\cite{Planck:2015fie,Planck:2018vyg} measurement 
of the CMB anisotropy is used to derive the 
$95\%$ confidence level (C.L.) upper limit 
on $\langle \sigma v \rangle$ for any given DM mass, 
following the methodology of~\cite{Slatyer:2015jla}. 
As shown in~\cite{Slatyer:2015jla}, the CMB 
constraints on the $\langle \sigma v \rangle - m_\chi$ 
plane of an annihilating WIMP DM is given by,
\begin{equation}
\epsilon_{\rm eff}(m_\chi)\frac{\langle \sigma v \rangle}{m_\chi} \lesssim 4.1 \times 10^{-28}\,{\rm cm}^3\,{\rm s}^{-1}\,{\rm GeV}^{-1},
\label{eq:CMBcons}
\end{equation}
where the weighted efficiency factor,
\begin{eqnarray}
\epsilon_{\rm eff}(m_\chi) &=& \frac{1}{2m_\chi}\int_0^{m_\chi}\,\underset{f \in {\rm SM}_1\overbar{\rm SM}_2}{\sum}\left(2\epsilon_{e^\pm}B_f\frac{dN_f}{dE_{e^\pm}} E_{e^\pm}dE_{e^\pm}+\epsilon_{\gamma}B_f\frac{dN_f}{dE_{\gamma}}E_{\gamma}dE_{\gamma}\right) .%\nonumber\\    
\end{eqnarray}
Here, $f$ represents a particular SM final state with 
branching ratio $B_f$ and the corresponding electron (photon) 
spectra are $dN_f/dE_{e^\pm}$ $\left(dN_f/dE_{\gamma}\right)$. 
The efficiency factors for $e^\pm$ and $\gamma$, i.e., 
$\epsilon_{e^\pm}$ and $\epsilon_{\gamma}$, are taken 
from~\cite{Slatyer:2015jla}. 

\begin{figure}[htb!]
\centering
\includegraphics[width=7.6cm,height=6.6cm]{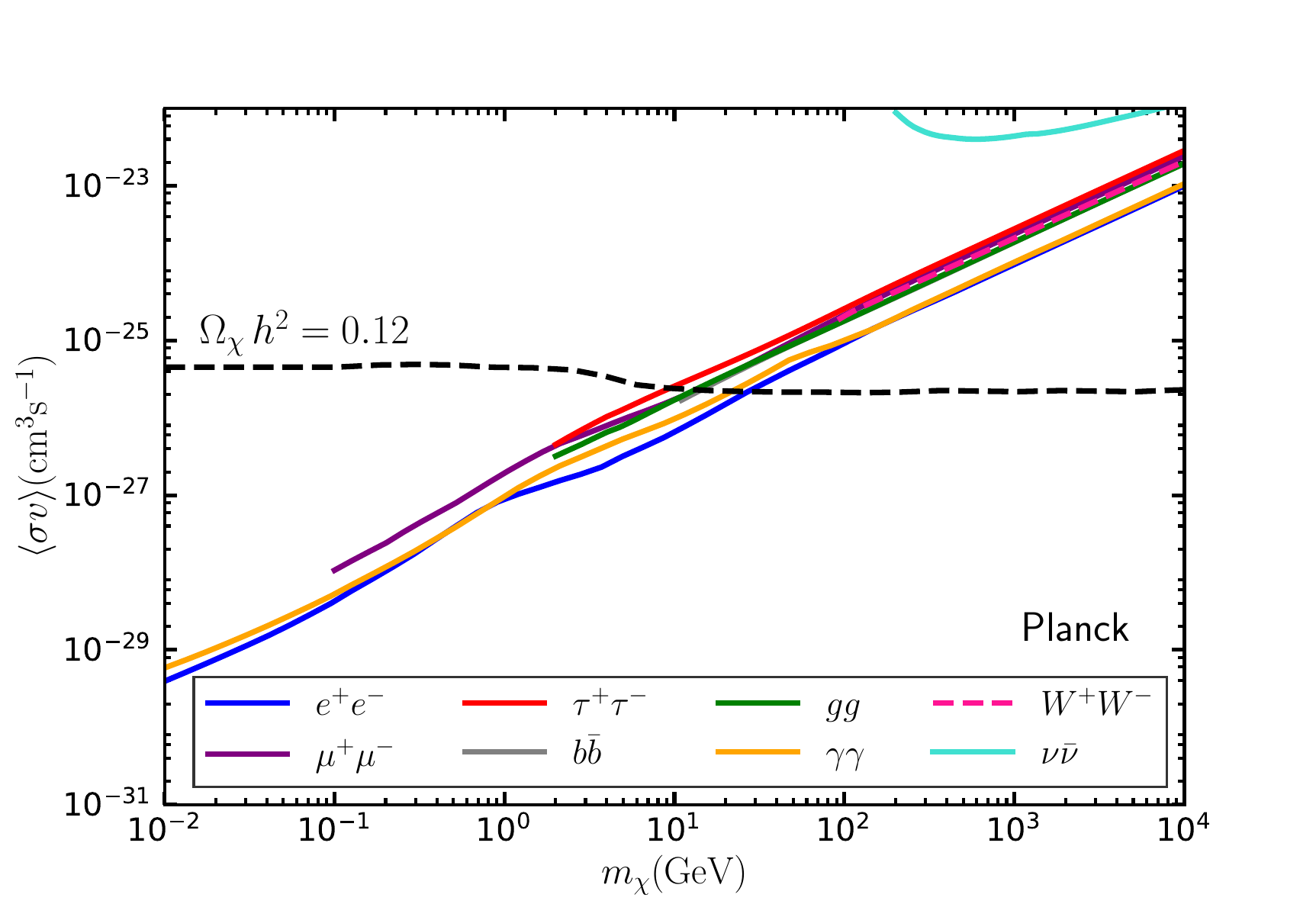}
\caption{The $95\%$ C.L. upper limits on $\langle \sigma v \rangle$, 
obtained from the Planck observation of the 
%temperature and polarization 
CMB anisotropies, are shown for eight 
representative annihilation channels, assuming $100\%$ 
BR for each individual final state.}
\label{fig:PlanckCMB}  
\end{figure} 

The $95\%$ C.L. upper limits on $\langle \sigma v \rangle$, 
are shown in Fig.~\ref{fig:PlanckCMB}, for eight representative 
annihilation channels, assuming $100\%$ BR for each channel. 
Throughout the considered $m_\chi$ range, the CMB constraints 
are strongest for $\gamma\gamma$ and $e^+e^-$, and weakest 
for $\nu\bar{\nu}$. In fact, for $m_\chi \lesssim 5\,{\rm GeV}$, 
the strongest constraints for the visible final states 
(assuming $s$-wave annihilations) are obtained from 
CMB~\cite{Cirelli:2020bpc}. 
% \red{In Ref.~\cite{Slatyer:2015jla}, the $e^-(e^+)$ and the 
% $\gamma$-ray spectra originating from DM annihilation induced 
% neutrino pairs had been obtained from~\cite{Cirelli:2010xx} 
% and the Planck data~\cite{Planck:2015fie} had been used. In 
% our analysis we have obtained the $e^-(e^+)$ and the $\gamma$ 
% spectra using 
% \texttt{Pythia}~\cite{Sjostrand:2014zea,Christiansen:2014kba}
% and used the most recent data of Planck~\cite{Planck:2018vyg} to 
% found that our limits are in agreement with those 
% presented in~\cite{Slatyer:2015jla}. 
We note that our limits agree with the limits 
obtained in~\cite{Slatyer:2015jla}, since we closely 
follow the analysis presented there.
%,in spite of the fact that 
% Ref.~\cite{Slatyer:2015jla} has obtained the DM annihilation 
% induced $e^-(e^+)$ and $\gamma$ spectra from~\cite{Cirelli:2010xx} 
% and used the data of~\cite{Planck:2015fie}, while we have 
% obtained the $e^-(e^+)$ and $\gamma$ spectra using 
% \texttt{Pythia}~\cite{Sjostrand:2014zea,Christiansen:2014kba} 
% and used the data of~\cite{Planck:2018vyg}. 
Moreover, we also checked that the limit for the 
$\nu\bar{\nu}$ final state is dominated by 
$\nu_e\bar{\nu}_e$, as shown in~\cite{Slatyer:2015jla}. 
The only uncertainties associated with the CMB 
limits stem from the $e^-(e^+)$ and the 
$\gamma$-ray spectra of DM annihilations which we 
have obtained using \texttt{Pythia}, and hence 
the CMB limits are free from all astrophysical 
uncertainties. 
%\sout{Note that, the uncertainties associated with the 
%$e^-(e^+)$ and the $\gamma$-ray spectra of DM annihilations
%not being large, CMB limits are quite robust~\cite{Leane:2018kjk}.}
% The theoretical uncertainties associated with these 
% limits come from the $e^-(e^+)$ and $\gamma$-ray 
% spectra of DM annihilations and 
% are not large (see~\cite{Leane:2018kjk}). 

\vspace{0.1cm}

\noindent
\subsection{Fermi-LAT} 

Dwarf spheroidal (dSph) galaxies are suitable targets 
to search for DM annihilation signals due to their 
high mass-to-light ratios and low astrophysical backgrounds. 
Hence, the $\gamma$-ray data in the energy range 
$500\,{\rm MeV} - 500\,{\rm GeV}$, obtained from the 
Fermi-LAT observation of dSphs~\cite{Fermi-LAT:2015att}, give 
strong constraints on the parameter space of 
annihilating WIMPs. 
%\sout{Fermi-LAT provides the likelihood curves 
%for each dSph, as a function of the integrated 
%energy flux:} 
Fermi-LAT collaboration~\cite{Fermi-LAT:2016uux} has 
performed a likelihood analysis 
of the six years of Fermi-LAT data observed from 
the directions of the dSphs and derived the 
$95\%$ C.L. upper limits on the 
$\langle \sigma v \rangle$ of WIMPs. They have 
provided the likelihood curve for each dSph as a 
function of the bin-by-bin integrated photon 
energy fluxes in the energy range $500\,{\rm MeV} - 500\,{\rm GeV}$~\cite{FermiLATdSphlike}. 
The integrated photon energy flux (in the energy bin $[E_{\rm min},E_{\rm max}]$) for the $i$-th dSph is given by:
\begin{equation}
\Phi_{E\,i} = \frac{\langle \sigma v \rangle\,J_i}{8\pi\,m^2_\chi}\int_{E_{\rm min}}^{E_{\rm max}} \underset{f \in {\rm SM}_1\overbar{\rm SM}_2}{\sum}B_f\frac{dN_f}{dE_\gamma}\,E_\gamma\,dE_\gamma,
\end{equation}
where $J_i$ is the J-factor of the $i$-th dSph. 
Following~\cite{Leane:2018kjk,Fermi-LAT:2015att}, 
we modify the likelihood curve for each dSph, 
$\mathcal{L}_i(\mu|\mathcal{D}_i)$~\cite{FermiLATdSphlike} 
(where $\mu$ is the model parameter and $\mathcal{D}_i$ is 
the $\gamma$-ray data for the $i$-th dSph), in the 
following way: 
\begin{eqnarray}
\tilde{\mathcal{L}}_i(\mu,J_i|\mathcal{D}_i) &=& \mathcal{L}_i(\mu|\mathcal{D}_i)%\nonumber\\&&\!\!\!
\times \dfrac{1}{\ln(10)\,J_i\,\sqrt{2\pi}\sigma_i}e^{-\left(\log_{10}(J_i)-\overbar{\log_{10}(J_i)}\right)^2/2\sigma^2_i}.%\nonumber\\
\label{eq:Fermilikes}
\end{eqnarray}
Here, $J_i$ acts as an additional nuisance parameter, 
%\sout{implying that the uncertainty associated with the 
%$J_i$ measurement is included. 
%to include the uncertainty associated with the 
%$J$-factor measurement of the $i$-th dSph. 
%In Eq.~\ref{eq:Fermilikes}}, 
with the values of $\overbar{\log_{10}(J_i)}$ 
and $\sigma_i$ %are 
obtained from~\cite{Fermi-LAT:2016uux}. 
The total likelihood function is obtained by 
multiplying the individual likelihoods of 
41 dSphs~\cite{Fermi-LAT:2016uux}, 
which include both kinematically confirmed and likely 
galaxies. This resulting function is then extremized 
(as described in~\cite{Leane:2018kjk}) to obtain the 
$95\%$ C.L. upper limit on $\langle \sigma v \rangle$.

\begin{figure}[htb!]
\centering
\includegraphics[width=7.6cm,height=6.6cm]{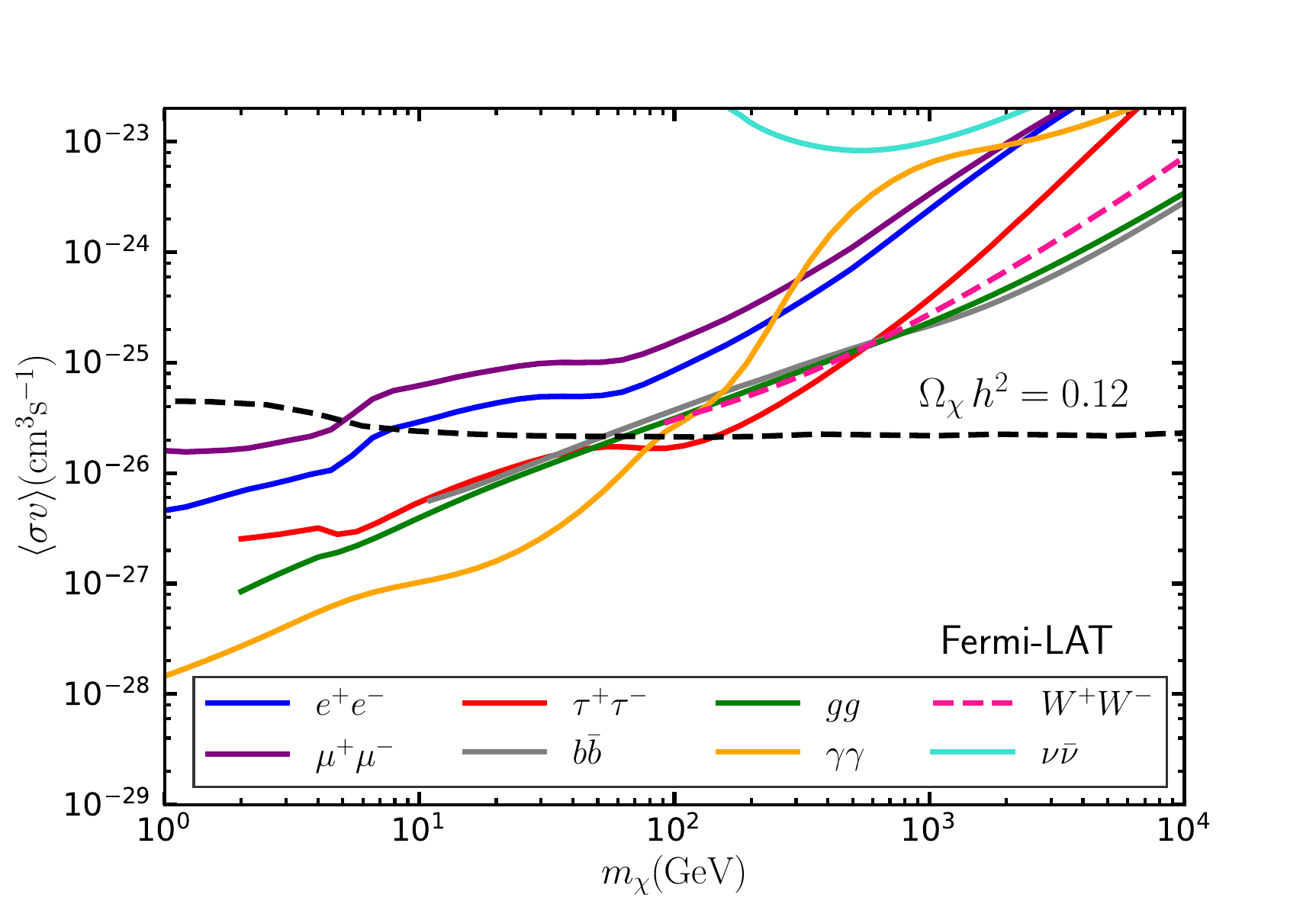}
\includegraphics[width=7.6cm,height=6.6cm]{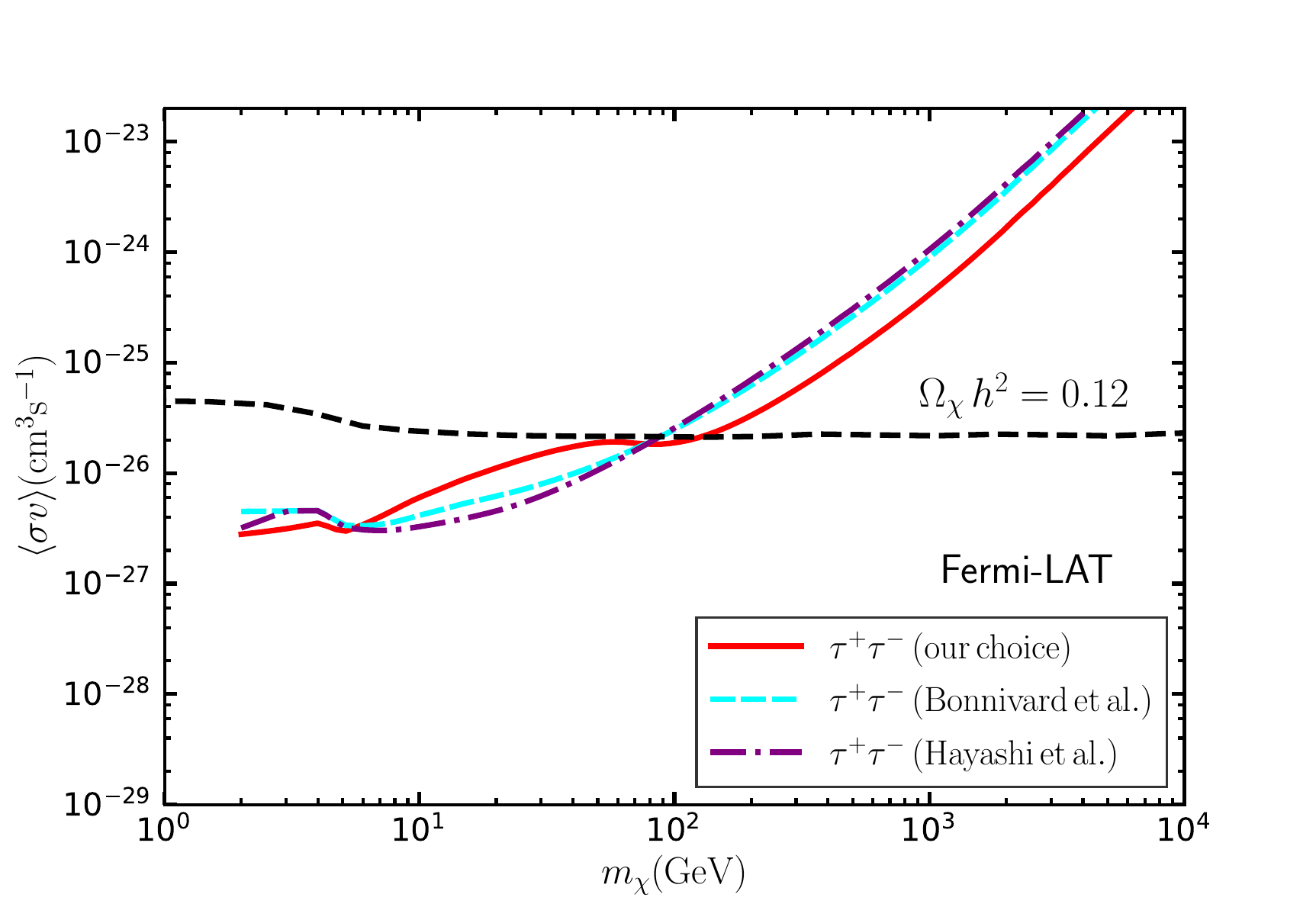}
\caption{{\it Left:} the $95\%$ C.L. upper limits on 
$\langle \sigma v \rangle$, obtained using the data of 
the Fermi-LAT dSph gamma-ray observation, are shown. 
Here, for illustrative purpose, the results for eight 
annihilation channels (assuming $100\%$ BR) are shown. 
{\it Right:} comparison of our limit for the 
$\tau^+\tau^-$ channel (red solid) with the limits obtained 
using the J-factors from~\cite{Bonnivard:2015xpq} (cyan dashed) 
and~\cite{Hayashi:2016kcy} (purple dashdotted).}
\label{fig:FermidSph}  
\end{figure} 

The resulting limits are shown in 
the left panel of Fig.~\ref{fig:FermidSph}. 
Here, we have shown the results for 
eight representative annihilation channels (each with $100\%$ BR). 
In this case, the $\gamma\gamma$ channel is most 
strongly constrained for $m_\chi \lesssim$ 
\textit{a few hundreds of GeV}, while for heavier DMs 
the constraints on the final states possessing hadronic 
decay modes (e.g., $\tau^+\tau^-$, $b\bar{b}$, $gg$) 
become stronger. Throughout the considered $m_\chi$ range, 
the constraints for the $\nu\bar{\nu}$ channel are the 
weakest. 
%\sout{Since the uncertainty associated with J-factors 
%are already taken into account, our results are quite robust.}

Note that the largest uncertainties associated with 
the Fermi-LAT $\gamma$-ray limits come from the measurements of 
the dSph J-factors. Here, we have considered the 
J-factors given in~\cite{Fermi-LAT:2016uux}. 
%In 
%Ref.~\cite{Fermi-LAT:2016uux}, the J-factors 
%for some dSphs are obtained} \red{\bf from the stellar 
%kinematic data~\cite{Geringer-Sameth:2014yza}, 
%while for the remaining dSphs the J-factors are %predicted using 
%the distance-scaling relation (see~\cite{Fermi-LAT:2016uux}). 
However, there exist several other measurements of the dSph J-factors (e.g., see~\cite{Bonnivard:2015xpq,Hayashi:2016kcy}). 
%those using optimized spherical Jeans 
%analysis of the surface brightness and the stellar kinematic 
%data~\cite{Bonnivard:2015xpq} and assuming non-spherical 
%dark halo models~\cite{Hayashi:2016kcy}. 
The upper limits on $\langle \sigma v \rangle$ obtained 
for DM annihilations to $\tau^+\tau^-$ considering the J-factors 
reported in~\cite{Bonnivard:2015xpq} (cyan dashed) 
and~\cite{Hayashi:2016kcy} (purple dashdotted) are compared 
against our limit (red solid) in the right panel of 
Fig.~\ref{fig:FermidSph}. The differences observed in the 
limits are attributed to the fact that J-factors for all 
41 dSphs considered in our analysis are not available 
in~\cite{Bonnivard:2015xpq,Hayashi:2016kcy} thereby leading 
to changes in the global likelihood for any given DM mass. 

% \blu{Finite $\nu\bar{\nu}$ BRs dilute the above limits, 
% unless the neutrinos carry energy 
% $\gtrsim 200\,{\rm GeV}$~\cite{Dutta:2022wuc}.} 
%The uncertainties associated with the 
%dSph J-factors~\cite{Leane:2018kjk} are taken 
%into account. 

\vspace{0.1cm}

%\item 
\noindent
\subsection{AMS-02} 
\label{sec:AMS02}
DM annihilations induced positrons undergo diffusion and 
energy losses while propagating through the 
Milky Way (MW) halo and can contribute to the AMS-02 
observation of the cosmic-ray positron flux in the energy 
range $500\,{\rm MeV} - 1\,{\rm TeV}$~\cite{AMS:2019rhg}.
The propagation equation 
of such positrons is~\cite{Leane:2018kjk}, 
\begin{eqnarray}
\frac{\partial N_i}{\partial t} &=& \vec{\nabla}.(D\vec{\nabla})N_i+\frac{\partial}{\partial p}(b(p,\vec{r}))N_i+Q_i(p,\vec{r})%\nonumber\\&&
+\sum_{j>i} \beta n_{gas}(\vec{r})\sigma_{ji}N_j-\beta n_{gas}(\vec{r})\sigma_i^{\rm in}(E_k)N_i,\nonumber\\
\label{eqn:difflossAMS}
\end{eqnarray}
where the source term for DM annihilations is given by,
\begin{equation}
Q_\chi(p,r,z) = \frac{\rho^2_\chi(r)\langle \sigma v \rangle}{2 m^2_\chi}\underset{f \in {\rm SM}_1\overbar{\rm SM}_2}{\sum} B_f \frac{dN_f}{dE_{e^\pm}}.    
\end{equation} 
In Eq.~\ref{eqn:difflossAMS}, we parameterize the 
MW diffusion parameter, $D(\rho,|\vec{r}|,z)$, 
as in~\cite{Leane:2018kjk}: 
\begin{equation}
D(\rho,|\vec{r}|,z) = D_0 e^{|z|/z_t} \left(\frac{\rho}{\rho_0}\right)^\delta,
\end{equation}
with reference rigidity $\rho_0 = 4\,$GV, diffusion coefficient 
$D_0 = 2.7\times 10^{28}\,{\rm cm}^2{\rm s}^{-1}$, 
diffusion index $\delta = 0.6$ and thickness of the 
axisymmetric diffusion zone $2z_t = 8\,{\rm kpc}$~\cite{Leane:2018kjk}. 
%\sout{Similar to DM decays~\cite{Dutta:2022wuc}, here, also, 
%other commonly used choices of $D(\rho,|\vec{r}|,z)$~\cite{Buch:2015iya,Genolini:2021doh} 
%do not affect the AMS-02 constraints significantly.} 
The DM induced positron fluxes depend on the values of the 
diffusion parameters. In order to illustrate how the diffusion 
parameters may affect the upper limits on $\langle \sigma v \rangle$, 
we shall consider other popular choices of 
$D(\rho,|\vec{r}|,z)$~\cite{Genolini:2019ewc,Genolini:2021doh}. 
%obtained from the AMS-02 data
%of secondary-to-primary ratios~\cite{Genolini:2019ewc,Genolini:2021doh}.}

On the other hand, the energy losses suffered by the 
positrons during their propagation (governed by the 
energy loss term $b(p,\vec{r})$), increases with 
the increase in the value of the galactic magnetic field, 
thereby weakening the corresponding limits. Therefore, 
in order to be conservative,
%Since the positrons lose more energy  while propagating 
%through a region of higher galactic magnetic field, 
we take somewhat larger value of the local magnetic field 
$B_\odot = 8.9\,\mu{\rm G}$~\cite{Leane:2018kjk}. 
For a smaller value of $B_\odot = 5.7\,\mu{\rm G}$~\cite{John:2021ugy,Beck:2008ty,Beck_2011}, %Beck:2013bxa}, 
our limits strengthen by less than a factor of two. 
The effect of solar modulation is %assumed to be well 
described by the force-field approximation 
with a modulation potential $\Phi$ = 0.6 GV~\cite{Cholis:2015gna}. 
On decreasing $\Phi$ to a value of 0.46 GV~\cite{Cholis:2015gna}, 
the constraints strengthen at most by a factor of two, 
for $m_\chi \lesssim 10\,{\rm GeV}$. 
The dependency of the derived constraints on the value of 
the solar modulation potential can be alleviated by utilising the 
$e^{\pm}$ data of Voyager 1 in the energy range 
$\sim 10-50\,$MeV~\cite{Stone2013Voyager1O}. However, the 
limits coming from the Voyager 1 data being 
weaker than the CMB limits (for $s$-wave 
annihilations)~\cite{Boudaud:2016mos}, 
our constraints on the total $\langle \sigma v \rangle$ 
will remain unaffected with the inclusion 
of Voyager 1 data.

We assume the DM distribution inside the MW halo, 
$\rho_\chi(r)$, follows the Navarro-Frenk-White (NFW)  
profile~\cite{Navarro:1995iw}: 
\begin{equation}
\rho_\chi(r) = \rho_s \left(\frac{r}{r_s}\right) \left(1+\frac{r}{r_s}\right)^{-2}, 
\label{eq:NFW}
\end{equation} 
with the scale radius $r_s = 20\,{\rm kpc}$, distance 
of the Sun from the galactic center (GC) 
$r_\odot = 8.5\,{\rm kpc}$ and local DM density 
$\rho_\odot = 0.25\,{\rm GeV}\,{\rm cm}^{-3}$~\cite{Leane:2018kjk}. 
Such a choice of $\rho_\odot$ is  
consistent~\cite{Salucci:2010qr,Bovy:2012tw,Benito:2019ngh,Benito:2020lgu} 
and yields most conservative upper limits on 
$\langle \sigma v \rangle$. %If one increases 
On raising $\rho_\odot$ to %a higher value of 
$0.7\,{\rm GeV}\,{\rm cm}^{-3}$~\cite{Salucci:2010qr,Bovy:2012tw,Benito:2019ngh,Benito:2020lgu}, 
the limits become stronger by a factor of eight. 

\begin{figure}[htb!]
\centering
\includegraphics[width=7.6cm,height=6.6cm]{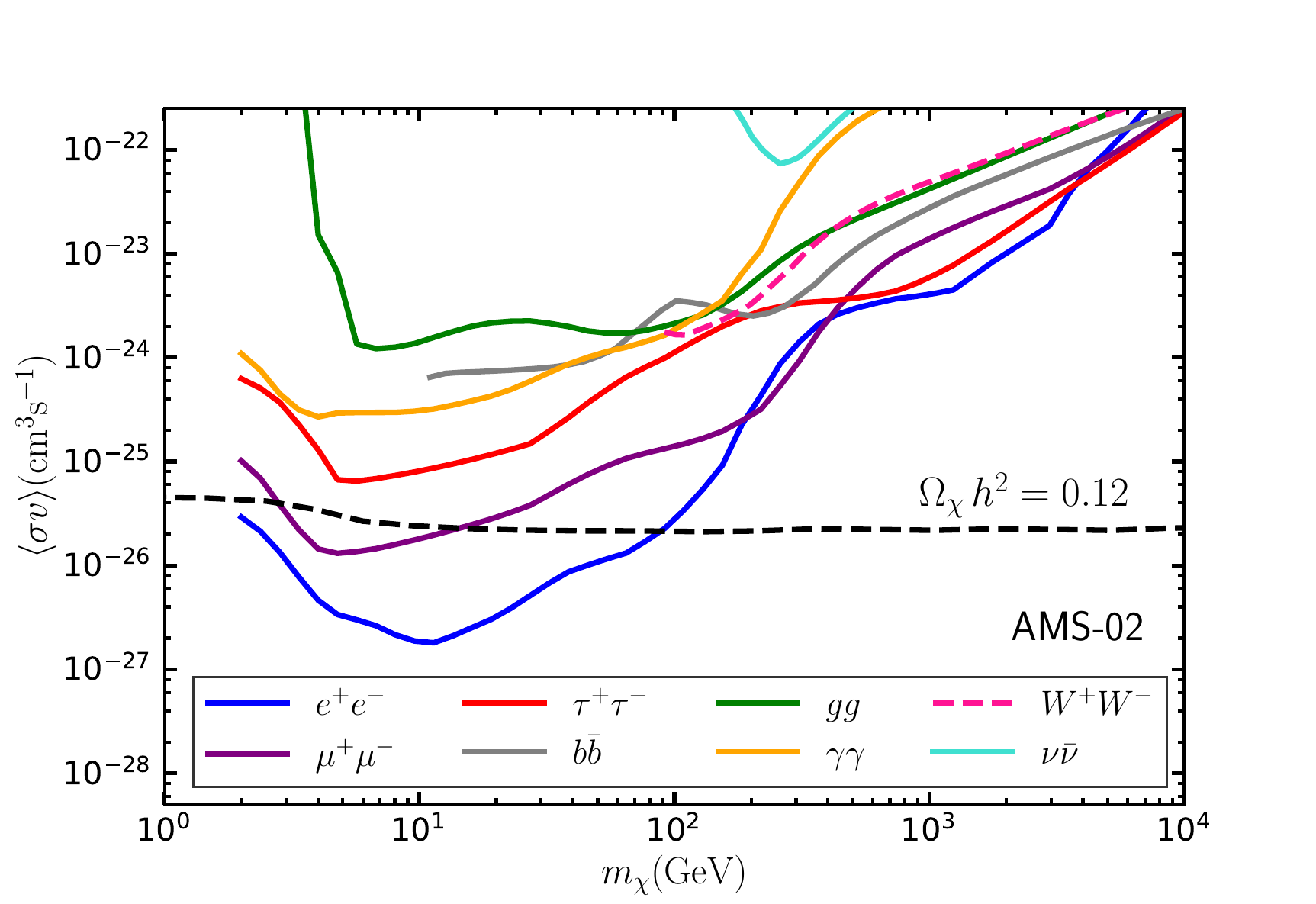}
\includegraphics[width=7.6cm,height=6.6cm]{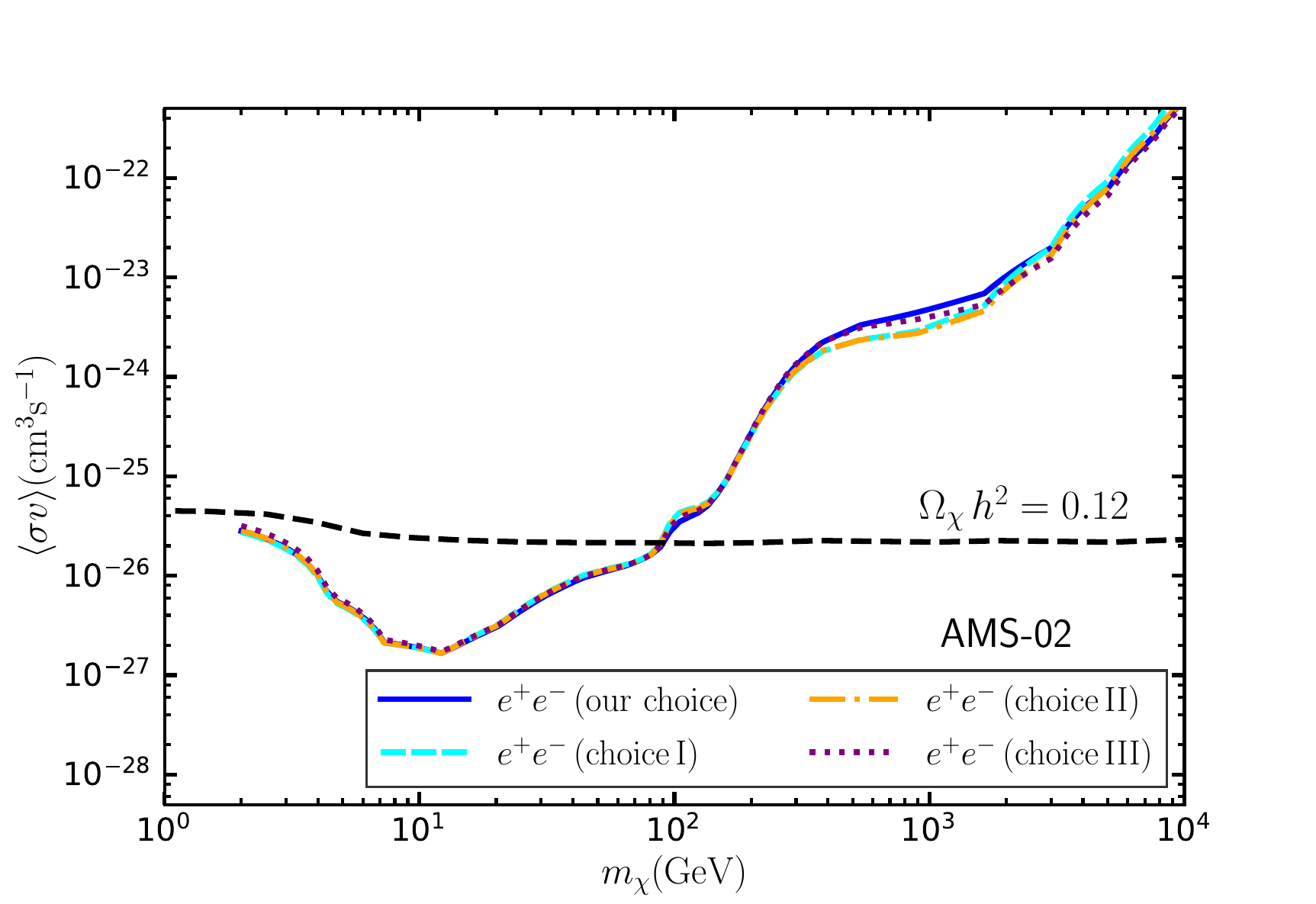}
\caption{ {\it Left:} the $95\%$ C.L. upper limits on 
$\langle \sigma v \rangle$, 
obtained using AMS-02 cosmic-ray positron flux data, 
are shown. Here, for the purpose of illustration, 
the results for eight different annihilation channels 
(assuming $100\%$ BR) are presented. 
{\it Right:} variation of the upper limits on 
$\langle \sigma v \rangle$ (for $e^+e^-$ channel) with 
the diffusion parameters is presented. Our limit is shown 
by the blue solid line, while the limits obtained using 
the diffusion parameter choices of~\cite{Genolini:2021doh} 
are shown by the cyan dashed, orange dashdotted and 
purple dotted lines.}
\label{fig:AMS02ep}  
\end{figure} 

% We followed the methodology 
% of~\cite{Leane:2018kjk,Dutta:2022wuc} to 
%\red{Method}
Following~\cite{Leane:2018kjk,Dutta:2022wuc}, we 
parameterize the AMS-02 measured positron flux spectrum 
as a polynomial function ($f(\alpha)$) of energy and 
define the $\chi^2$ as
\begin{equation}
\chi^2 = \sum_{i = bins} \frac{(f_i(\alpha)-\mathcal{D}_i)^2}{\sigma^2_i},   
\end{equation}
where $\mathcal{D}_i$ represents the AMS-02 measured 
positron flux data in the $i$-th energy bin and 
$\sigma_i$ is the associated 
uncertainty~\cite{AMS:2019rhg}. This $\chi^2$ is 
extremized to obtain the best-fit 
$\chi^2$, i.e., $\chi^2_{\rm bf}$, which corresponds to 
the best-fit values of $\{\alpha\}$. Then we add the DM 
induced signal to $f(\alpha)$ and vary $\{\alpha\}$ within 
$30\%$ of their best-fit values. To obtain the AMS-02 
constraints on $\langle \sigma v \rangle$ (at $95\%$ C.L.) 
for any given $m_\chi$, we slowly increase 
$\langle \sigma v \rangle$ until 
$\chi^2 = \chi^2_{\rm bf} + 2.71$. 
The resulting limits for eight 
representative annihilation channels are 
shown in the left panel of 
Fig.~\ref{fig:AMS02ep}, 
assuming $100\%$ BR for each channel. Here, $e^+e^-$ final 
state is most strongly constrained while $\nu\bar{\nu}$ 
is most weakly constrained. 

In the right panel of Fig.~\ref{fig:AMS02ep}, we have shown 
the variation of the upper limit on $\langle \sigma v \rangle$ with the 
choice of the diffusion parameters assuming $100\%$ BR for $e^+e^-$ 
channel. In addition to our choice of $D(\rho,|\vec{r}|,z)$ 
(blue solid line), we have also considered three other possibilities, 
consistent with the current AMS-02 data~\cite{Genolini:2021doh}, 
namely, $D_0 = 2\times 10^{28}\,{\rm cm}^2{\rm s}^{-1}$, 
$\delta = 0.49$, $z_t = 8.4\,{\rm kpc}$ (choice I; cyan dashed line), 
$D_0 = 1.1\times 10^{28}\,{\rm cm}^2{\rm s}^{-1}$, $\delta = 0.499$, 
$z_t = 4.67\,{\rm kpc}$ (choice II; orange dashdotted line) 
and $D_0 = 5\times 10^{27}\,{\rm cm}^2{\rm s}^{-1}$, 
$\delta = 0.509$, $z_t = 2.56\,{\rm kpc}$ (choice III; purple 
dotted line), each with reference rigidity $\rho_0 = 1\,$GV. 
This clearly demonstrates that our limits are not very 
sensitive to the choice of the diffusion parameters. 

\vspace{0.1cm}

\noindent
\subsection{H.E.S.S}

H.E.S.S is quite sensitive to very high energy 
gamma-rays. Thus, for $m_\chi \gtrsim 200\,{\rm GeV}$, 
the constraints obtained from the H.E.S.S gamma-ray 
observation from the Galactic halo~\cite{HESS:2022ygk} 
are stronger than those coming from other data. 
We assume that a NFW profile (with $r_s = 20\,{\rm kpc}$, 
$r_\odot = 8.5\,{\rm kpc}$ and 
$\rho_\odot = 0.25\,{\rm GeV}\,{\rm cm}^{-3}$) 
describes the MW DM 
distribution, %of the MW halo, 
and adopt the ON-OFF procedure, as described 
in~\cite{HESS:2022ygk}, to derive the H.E.S.S 
constraints. The ON and the OFF 
source regions are two well-separated regions 
inside the galactic halo, such that, the DM 
induced signal events coming from the ON region 
are significantly larger than 
those coming from the OFF region. 
Here, ON region is an annular ring centered 
around the GC with inner radius $0.5^\circ$ 
and outer radius $3^\circ$, excluding the 
galactic latitude $|b| < 0.3^\circ$ region 
along the galactic plane. This ON region is further 
subdivided into 25 regions of interests (ROIs), each 
being an annular ring of width $0.1^\circ$\cite{HESS:2022ygk}. 
OFF source regions are of the same size as 
the ON region, but ON and OFF regions are situated 
symmetrically about the pointing positions of the 
telescope~\cite{HESS:2022ygk}. 

\begin{figure}[htb!]
\centering
\includegraphics[width=7.6cm,height=6.6cm]{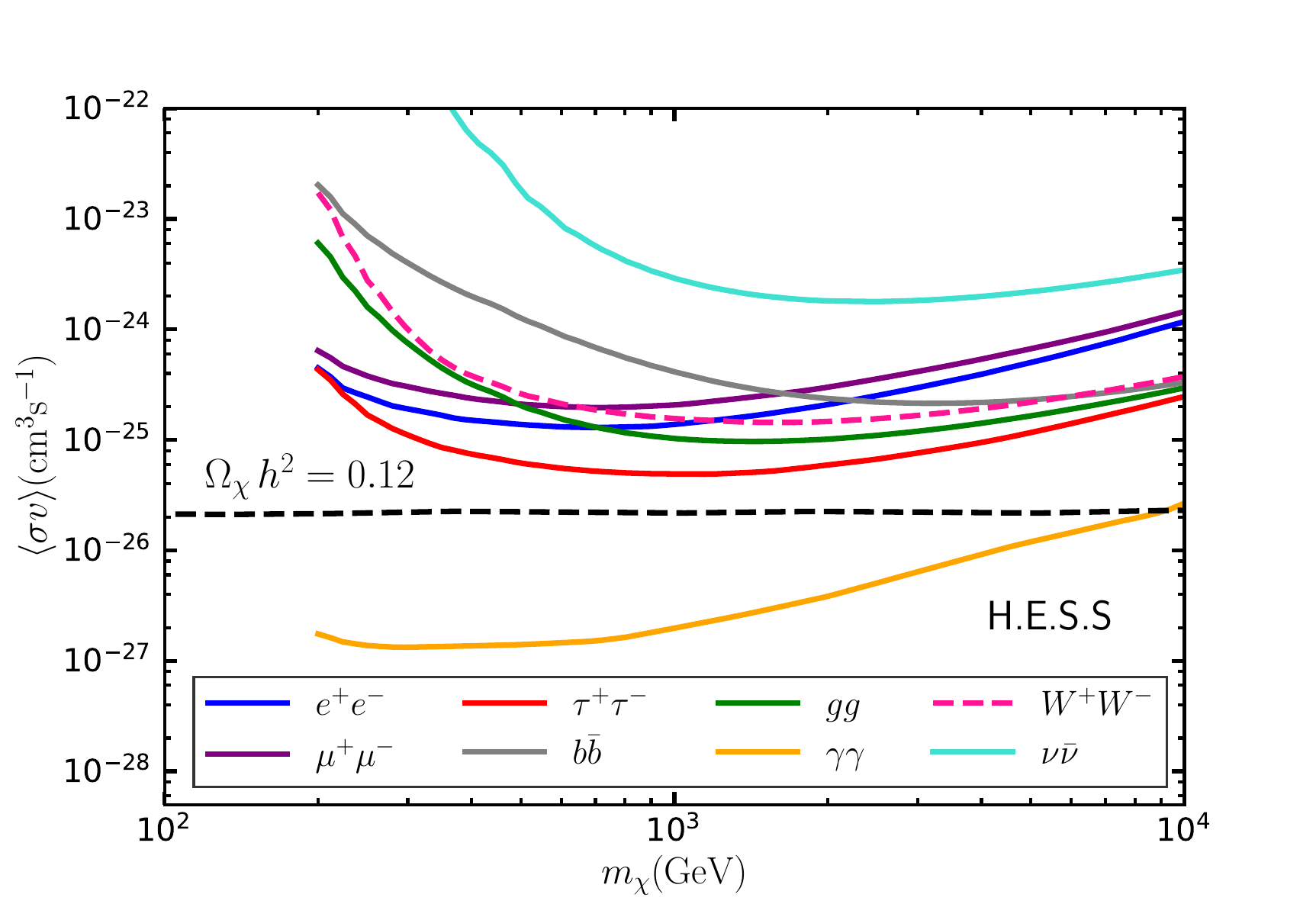}
\caption{For eight representative annihilation channels, the 
$95\%$ C.L. upper limits on $\langle \sigma v \rangle$, 
obtained from the Galactic halo gamma-ray observation by H.E.S.S, are 
shown. Here, we have assumed $100\%$ BR for each individual 
SM final state.}
\label{fig:HESSGC}  
\end{figure} 

The number of signal events expected from the ON and 
the OFF regions are given by~\cite{HESS:2022ygk}:
\begin{eqnarray}
N^S_{i,{\rm ON(OFF)}} = \frac{\langle \sigma v \rangle\,J_{i,{\rm ON(OFF)}}}{8\pi\,m^2_\chi}T_{\rm obs}\int_{E_{\rm th}}^{m_\chi}\int_{0}^{\infty}dE^\prime_\gamma dE_\gamma \underset{f \in {\rm SM}_1\overbar{\rm SM}_2}{\sum}B_f\frac{dN_f}{dE_\gamma}(E_\gamma)\,A_{\rm eff}(E_\gamma)\,R(E_\gamma, E^\prime_\gamma),\nonumber\\
\label{eq:HESSsig}
\end{eqnarray}
where $i$ denotes the $i$-th ROI and $J_{i,{\rm ON(OFF)}}$ 
represent the $J$-factors for the $i$-th ON (OFF) regions. 
In Eq.~\ref{eq:HESSsig}, $A_{\rm eff}$ represents the 
H.E.S.S effective area~\cite{Veh:2018ckl}, 
$R(E_\gamma, E^\prime_\gamma)$ 
is a gaussian function accounting for the detector energy 
resolution of $10\%$~\cite{HESS:2022ygk} and $T_{\rm obs}$ 
is the H.E.S.S observation time which is 
546 hours~\cite{HESS:2022ygk}. Taking the expected 
number of background events ($N^B_i$) into account, we 
construct a likelihood function~\cite{HESS:2022ygk},
%\blu{Using the H.E.S.S effective area~\cite{Veh:2018ckl} 
%and an energy resolution of $10\%$~\cite{HESS:2022ygk}, we 
%calculate both the number of signal as well as background 
%events expected from the ON and the OFF source regions, 
%assuming 546 hours of observation time~\cite{HESS:2022ygk} 
% , which is combined with the 
% observed data to obtain the $95\%$ C.L. upper limits 
% on $\langle \sigma v \rangle$, following~\cite{Cowan:2010js}. 
%and construct a likelihood function,}
\begin{eqnarray}
\mathcal{L}_i =\dfrac{(N^S_{i,{\rm ON}}+N^B_i)^{N_{{\rm ON},i}}}{N_{{\rm ON},i} !}e^{-(N^S_{i,{\rm ON}}+N^B_i)}%\nonumber\\& \times 
\dfrac{(N^{S}_{i,{\rm OFF}}+N^B_i)^{N_{{\rm OFF},i}}}{N_{{\rm OFF},i} !}e^{-(N^{S}_{i,{\rm OFF}}+N^B_i)},    
\end{eqnarray}
where $N_{{\rm ON},i}$ and $N_{{\rm OFF},i}$ represent the 
respective number of observed $\gamma$-ray events from the 
$i$-th ON and OFF regions. 
%$N^S_{i,{\rm ON}}$ and $N^{S}_{i,{\rm OFF}}$ 
%signify the number of signal events expected from the $i$-th ON 
%and OFF regions, respectively, and $N^B_i$ represents the predicted 
%number of background events from those regions.} 
The total likelihood function, 
$\mathcal{L}=\underset{i}{\prod} \mathcal{L}_i$, is then 
extremized following the methodology of~\cite{Cowan:2010js}, 
to obtain the $95\%$ C.L. upper limits on 
$\langle \sigma v \rangle$.

%These predicted events 
%are combined with the observed data~\cite{HESS:2022ygk} 
%%and a likelihood-ratio test is performed, 
%following the methodology of~\cite{Cowan:2010js}, 
%to obtain the $95\%$ C.L. upper limits on 
%$\langle \sigma v \rangle$. 
% We combine these events with the observed 
% data~\cite{HESS:2022ygk} and perform a 
% likelihood-ratio test, following~\cite{Cowan:2010js}, 
% to obtain the $95\%$ C.L. upper limits on 
% $\langle \sigma v \rangle$. 

The resulting limits for eight representative annihilation 
channels, each with $100\%$ BR, are shown in Fig.~\ref{fig:HESSGC}. 
In this case, $\gamma\gamma$ is most strongly 
constrained while $\nu\bar{\nu}$ is most weakly 
constrained. The largest possible uncertainty in 
this case is associated with the value of the 
$\rho_\odot$ and its variation affects 
these constraints at most by a factor of eight.

Since $e^-(e^+)$ and $\gamma$-ray fluxes 
from the DM annihilation induced primary 
$\nu\bar{\nu}$ pairs are produced via the 
radiation of electroweak gauge bosons, $\nu\bar{\nu}$ 
channel is most weakly constrained, especially for 
low $m_\chi$. Nevertheless, $W$-radiations from these 
neutrinos %\sout{cause a difference in} 
are important in the context of H.E.S.S observation 
%contribute to H.E.S.S constraints 
for $m_\chi \gtrsim 800\,{\rm GeV}$. 

%\vspace{1mm}
%\noindent
\section{Limits from neutrino observations}
\label{sec:sec3}
Neutrino telescopes detect the (anti)neutrinos 
(instead of the $e^-(e^+)$ and $\gamma$-ray photons) 
coming from DM annihilations, 
thereby providing strongest constraints for 
the primary $\nu\bar{\nu}$ channel. 
%\sout{Hence, the resulting constraints are strongest 
%for the primary $\nu\bar{\nu}$ final state.} 
Here, we consider the data of three different neutrino 
observations, assuming the DM distribution inside 
the MW halo follows a NFW profile (see Eq.~\ref{eq:NFW}) 
with the conservative set of parameters, discussed earlier.

\vspace{0.1cm}

\noindent
\subsection{Super-Kamiokande}

\noindent
\underline{\textbf{Atmospheric neutrino observation:}}
The neutrino fluxes measured from the MW halo by the 
Super-Kamiokande~\cite{Super-Kamiokande:2015qek} 
is used to constrain 100 MeV - 10 TeV 
thermal WIMPs~\cite{Primulando:2017kxf,Super-Kamiokande:2020sgt}.
In obtaining these limits, the ON-OFF analysis method 
outlined in~\cite{Super-Kamiokande:2020sgt} is adopted, 
assuming the ON region to be a circle centered 
around the GC and the OFF region to be another 
equal-sized circular region offset by $180^\circ$ 
in right ascension with respect to the 
GC~\cite{Super-Kamiokande:2020sgt}. 
DM annihilation induced neutrino+antineutrino flux 
distributions from the ON and the OFF source regions are 
given by:
\begin{eqnarray}
\frac{d\Phi_{\rm ON(OFF)}}{dE_\nu} = \frac{\langle \sigma v \rangle\,J_{\rm ON(OFF)}}{8\pi\,m^2_\chi}\underset{f \in {\rm SM}_1\overbar{\rm SM}_2}{\sum}B_f\left(\frac{dN_f}{dE_\nu}+\frac{dN_f}{dE_{\bar{\nu}}}\right),    
\end{eqnarray}
where, $J_{\rm ON}$ and $J_{\rm OFF}$ represent the 
$J$-factors for the ON and the OFF regions, respectively. 
While obtaining the neutrino (antineutrino) spectrum 
$dN_f/dE_{\nu}$ ($dN_f/dE_{\bar{\nu}}$), the effects of 
neutrino oscillations during propagation has been 
taken into account, following~\cite{Super-Kamiokande:2020sgt}.
Using these flux distributions and the formulae given 
in~\cite{Covi:2009xn}, we obtain the number of signal 
events expected %and background 
from ON and OFF source regions. %are calculated, 
%including the effects of neutrino 
%oscillations~\cite{Super-Kamiokande:2020sgt}. 
For the corresponding background events, the 
distributions of  atmospheric neutrinos are 
obtained from~\cite{Honda:2011nf}. 
For $m_\chi \lesssim 1\,{\rm GeV}$, the fully 
contained (FC) sub-GeV data sample collected 
from source regions of half-opening angle 
$60^\circ$~\cite{Super-Kamiokande:2020sgt} is used, 
while, for $m_\chi \gtrsim 1\,{\rm GeV}$, the 
partially contained (PC) and Upward-going through 
muon (UP-$\mu$) events are considered assuming 
source regions of half-opening angles $20^\circ$ 
and $10^\circ$, respectively~\cite{Super-Kamiokande:2020sgt}.

\begin{figure}[htb!]
\centering
\includegraphics[width=7.6cm,height=6.6cm]{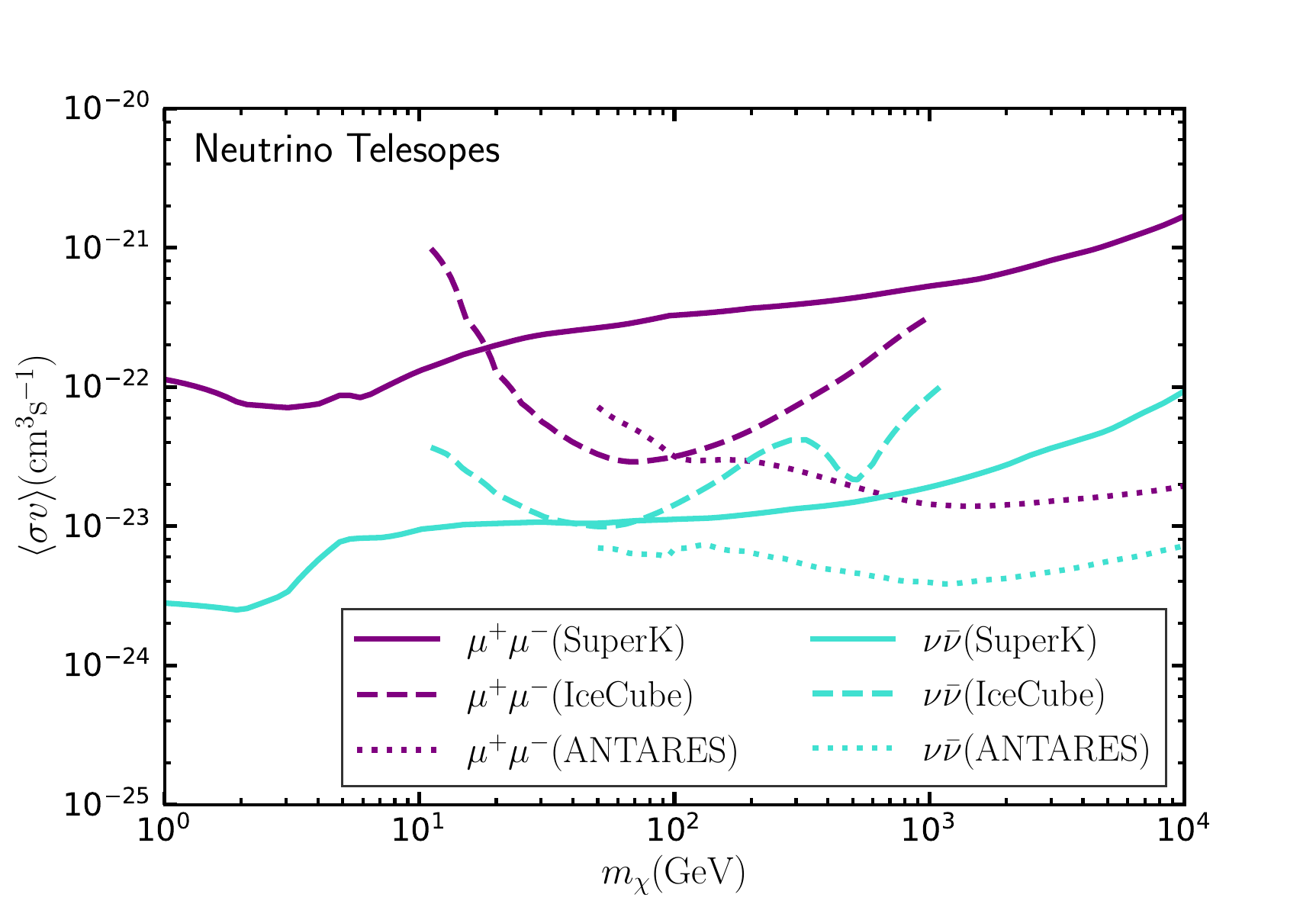}
\caption{$95\%$ C.L. upper limits on 
$\langle \sigma v \rangle$ for $\nu\bar{\nu}$ (cyan) 
and $\mu^+\mu^-$ (purple) final states, obtained using 
the data of Super-Kamiokande (solid lines), 
IceCube (dashed lines) and ANTARES (dotted lines) 
telescopes are shown, assuming $100\%$ BR for each channel.
}
\label{fig:nuobs}
\end{figure}

For each event category, we calculate the asymmetry, 
$A=(N_{\rm ON}-N_{\rm OFF})/(N_{\rm ON}+N_{\rm OFF})$, 
with $N_{\rm ON}$ and $N_{\rm OFF}$ being the total number 
of signal+background events expected from 
the ON and the OFF source regions, respectively. 
This asymmetry parameter is compared with 
its measured value (provided in~\cite{Super-Kamiokande:2020sgt}) 
to obtain the upper limit on $\langle \sigma v \rangle$ 
for any given $m_\chi$. Assuming $100\%$ BR for each 
channel, one finds that $\nu\bar{\nu}$ (cyan solid line in 
Fig.~\ref{fig:nuobs}) is the most strongly constrained 
final state, while $\mu^{+}\mu^{-}$ (purple solid line 
in Fig.~\ref{fig:nuobs}) is the second most strongly 
constrained channel. The limits for the remaining 
annihilation channels are weaker. With the variation 
of $\rho_\odot$, these limits may vary by a factor 
of eight.

\noindent
\underline{\textbf{Diffuse Supernova Neutrino Background (DSNB) observation:}}
For 10 MeV - 100 MeV WIMPs, the strongest 
constraint for annihilation into the $\nu\bar{\nu}$ 
channel comes from the data of the low-energy neutrino 
search by the Super-Kamiokande which is commonly used 
to constrain the diffuse supernova neutrino 
background (DSNB)~\cite{Super-Kamiokande:2021jaq}. 
5823 live time days of all-sky data, ranging from 
9 MeV - 88 MeV positron energies~\cite{Super-Kamiokande:2021jaq}, 
have been used (following the methodology 
of~\cite{PhysRevD.97.075039,Palomares-Ruiz:2007trf}) 
%Olivares-DelCampo:2017feq
to derive the $95\%$ C.L. upper limits on $\langle \sigma v \rangle$. 
In this case, the background positron distributions 
are obtained from~\cite{Super-Kamiokande:2011lwo}.
%\red{Method}

\vspace{0.1cm}

\noindent
\subsection{IceCube} 

Data from 1005 days of neutrino observation 
of the MW halo by the IceCube is used to 
constrain 10 GeV - 1 TeV WIMPs~\cite{IceCube:2017rdn}. 
%following the methodology of~\cite{IceCube:2017rdn}. 
Assuming a source region, extended from 0 to $2\pi$ 
radian in right ascension (RA) and -1 to +1 in 
declination (DEC)~\cite{IceCube:2017rdn}, 
the distribution of the DM induced signal events, 
\begin{eqnarray}
N^S_i = \frac{\langle \sigma v \rangle\,J_{i}}{8\pi\,m^2_\chi}T_{\rm obs}\,\int_0^{m_\chi}\underset{f \in {\rm SM}_1\overbar{\rm SM}_2}{\sum}
B_f\,\left(\frac{dN_f}{dE_\nu}\,A_{\rm eff}(E_\nu)\,dE_\nu + \frac{dN_f}{dE_{\bar{\nu}}}\,A_{\rm eff} (E_{\bar{\nu}})\,dE_{\bar{\nu}}\right),
\label{eq:ICsig}
\end{eqnarray}
is obtained as a function of 
RA and DEC~\cite{IceCube:2017rdn}, including 
the effects of neutrino 
oscillations~\cite{Super-Kamiokande:2020sgt}. 
In Eq.~\ref{eq:ICsig}, $J_i$ represents the $J$-factor 
for the $i$-th spatial bin, $T_{\rm obs}$ is the IceCube 
observation time and $A_{\rm eff}$ is the 
IceCube effective area~\cite{IceCube:2017rdn}. 
The distribution in Eq.~\ref{eq:ICsig} is normalized to 
obtain the normalized signal event distribution $f_s$, 
while, the normalized background distribution ($f_B$)
for the considered observation region is obtained 
from~\cite{IceCube:2017rdn}. 
%and the 
%IceCube effective areas for different neutrino flavours 
%\sout{and the normalized background distribution 
%for the considered source region,} 
%are obtained from~\cite{IceCube:2017rdn}.
%The likelihood extremization technique 
%of~\cite{Feldman:1997qc} is then employed 
%to derive the $95\%$ C.L. upper limit on 
%$\langle \sigma v \rangle$ for any given DM mass. 

These normalized signal and background distributions are 
combined to form the total event distribution~\cite{IceCube:2017rdn,ANTARES:2020leh}, 
\begin{equation}
f(\mu) = \mu\,f_s + (1-\mu)\,f_B,
\label{eq:combev}
\end{equation}
where $\mu \in [0,1]$ represents the fraction of 
signal events present in the total sample. Thereafter, 
$f(\mu)$ is combined with the observed data to construct 
a likelihood function (as in~\cite{IceCube:2017rdn,ANTARES:2020leh}):
\begin{equation}
\mathcal{L}(\mu) = \underset{i}{\prod} \dfrac{(n^{\rm tot}_{\rm obs}f_i(\mu))^{n_{{\rm obs},i}}}{n_{{\rm obs},i}!}\,e^{-n^{\rm tot}_{\rm obs}f_i(\mu)},
\label{eq:likeIC}
\end{equation}
with $n^{\rm tot}_{\rm obs}$ being the total number of 
observed neutrino events and $n_{{\rm obs},i}$ 
representing the neutrino events observed in the $i$-th 
spatial bin by the IceCube. This likelihood function 
is then extremized following the methodology %discussed in
of~\cite{Feldman:1997qc} to derive the $95\%$ C.L. 
upper limit on $\mu$, i.e., $\mu_{95\%}$, for each 
annihilation channel and a given DM mass. This upper limit 
is then converted to $95\%$ C.L. upper limit on 
$\langle \sigma v \rangle$. We show the results for 
$\nu\bar{\nu}$ (cyan dashed line) and $\mu^{+}\mu^{-}$ 
(purple dashed line) in Fig.~\ref{fig:nuobs}, 
assuming $100\%$ BR for each channel.

\vspace{0.1cm}

\noindent
\subsection{ANTARES} 

Using 2102 days of data of the neutrino observation 
towards the central region of the MW halo by the ANTARES 
telescope~\cite{Albert:2016emp}, we constrain 
WIMPs of masses 50 GeV - 100 TeV. 
% \sout{Here we have considered only the $\nu_\mu$ 
% events~\cite{Albert:2016emp}.
The source region is assumed to be a circle 
centered around the GC with half-opening angle 
$\psi = 30^\circ$~\cite{Albert:2016emp} and  
the normalized angular distribution of the 
$\nu_\mu$-type signal events ($f_s$) is 
calculated as a function of $\psi$ (as in the 
case of IceCube). Here, the detector effective 
area is taken from~\cite{ANTARES:2017dda}. The normalized 
background event distribution ($f_B$) for the same 
observation region is obtained from~\cite{Albert:2016emp}. 
%The detector effective area has been taken 
%from~\cite{ANTARES:2017dda}. 
% Following the same methodology as in the case of IceCube, 
% here, too, we perform a likelihood extremization to 
% obtain the $95\%$ C.L. upper limit on 
% $\langle \sigma v \rangle$ for any given $m_\chi$.} 
%Taking the detector effective area 
%from~\cite{ANTARES:2017dda} 
Then, using the neutrino data observed by ANTARES, 
predicted signal distribution ($f_s$) and background 
distribution ($f_B$), we construct a likelihood function 
as in Eq.~\ref{eq:likeIC} and follow the methodology 
of~\cite{Albert:2016emp,ANTARES:2020leh} to obtain the 
$95\%$ C.L. upper limit on the fraction of signal 
events $\mu$, i.e., $\mu_{95\%}$. This is then converted 
to $95\%$ C.L. upper limit on $\langle \sigma v \rangle$, 
for any given $m_\chi$. The resulting constraints for 
$\nu\bar{\nu}$ (cyan dotted line) and $\mu^{+}\mu^{-}$ 
(purple dotted line) channels are presented in 
Fig.~\ref{fig:nuobs}.

Note that, for $m_\chi \gtrsim 800\,{\rm GeV}$, 
H.E.S.S provides the strongest constraints for 
$\nu\bar{\nu}$ (compare Figs.~\ref{fig:HESSGC} 
and \ref{fig:nuobs}).

%\vspace{1mm}
%\noindent
\section{Constraints on total annihilation cross-section} %\\
\label{sec:sec4}

%\noindent
The constraints coming from the data of the indirect 
search observations depend on the spectrum of photons, 
$e^-(e^+)$ and neutrinos produced in DM annihilations. 
As a result, the BRs of different annihilation channels play 
important roles in constraining total $\langle \sigma v \rangle$. 
Ref.~\cite{Leane:2018kjk} provides the BR-independent 
upper limit on the total $\langle \sigma v \rangle$ of a 
thermal WIMP without taking into account the 
%contribution of the 
possibility of annihilation into $\nu\bar{\nu}$ channel 
and the data of the neutrino observations. 
Although Ref.~\cite{Slatyer:2015jla} 
has pointed out that 
the CMB limits for the neutrino final states 
are comparable to the limits coming from the 
neutrino 
observations~\cite{IceCube:2013bas,IceCube:2014rqf,IceCube:2015rnn}, 
a dedicated analysis considering all possible 
BRs of the $\nu\bar{\nu}$ channel and using 
the updated data from the 
neutrino observations is still lacking. 
%\sout{However, as 
%shown in sec.~\ref{sec:sec3}, the constraints %coming from the 
%neutrino telescopes may also be relevant.} 
In addition, the inclusion of H.E.S.S data 
(primarily from the gamma-ray observation of 
the Galactic halo ~\cite{HESS:2022ygk}) also 
has important implications for annihilation of 
multi-TeV WIMPs. Here, we consider the 
possibilities of DM annihilations 
into all possible SM particle pairs, including 
$\nu\bar{\nu}$, 
and use the data of all observations 
(discussed in secs.~\ref{sec:sec2} 
and~\ref{sec:sec3}) to obtain the most 
general BR-independent upper limit on 
the total $\langle \sigma v \rangle$. 

The method for obtaining this BR-independent 
upper limit on the total $\langle \sigma v \rangle$ 
is broadly as follows:
%\sout{The BR-independent upper limit on the total 
%$\langle \sigma v \rangle$ of a generic 
%thermal WIMP is obtained as follows:}
 
\noindent
\begin{itemize}
\item For each $m_\chi$, in the range 
$10\,{\rm MeV} - 100\,{\rm TeV}$, 
we scan over all possible BR combinations of the 
kinematically allowed two-body annihilation channels, 
with grid-size of %not exceeding 
$2\%$ along any BR axis, ensuring that the BRs 
of all channels always add up to $100\%$. 

\item For each such combination, we derive the 
$95\%$ C.L. upper limits on $\langle \sigma v \rangle$ 
coming from all observations, and find the smallest 
(i.e., the strongest) limit among these. This limit 
represents the maximum allowed %\sout{value of the} 
$\langle \sigma v \rangle$, consistent with all 
observational data, for that particular BR combination. 

\item Among these allowed values of 
$\langle \sigma v \rangle$ obtained for 
different BR combinations, we find the largest 
(i.e., the weakest) one, which is the maximum 
allowed %\sout{value of the} 
total $\langle \sigma v \rangle$, 
for the considered value of $m_\chi$. 
%The corresponding BR combination 
%is known as the threshold BR combination. 
\end{itemize}

For further details of the analysis methodology, 
see~\cite{Leane:2018kjk,Dutta:2022wuc}. 
In obtaining the limits on %total 
$\langle \sigma v \rangle$, we assume a single-component 
WIMP accounts for the entire DM content. %in the galaxies. 

\begin{figure}[htb!]
\centering
\includegraphics[width=7.6cm,height=6.6cm]{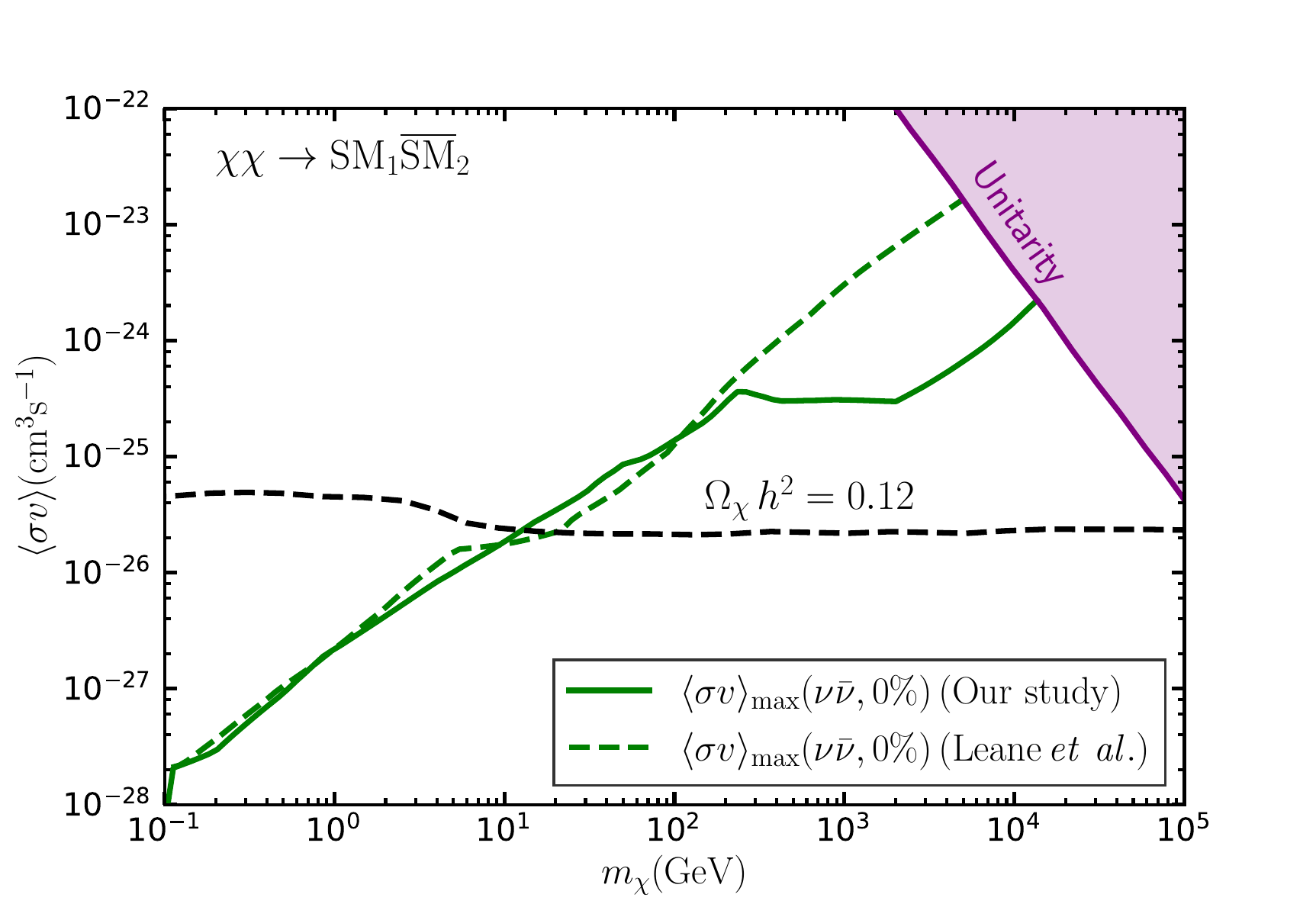}
\caption{For DM annihilations into visible 
SM final states only, the $95\%$ C.L. upper 
limit on total $\langle \sigma v \rangle$ 
for the $m_\chi$ range 
$100\,{\rm MeV} - 100\,{\rm TeV}$ (green solid line) 
is shown. Due to the inclusion of H.E.S.S 
%\sout{and other updated observational} 
data, our limit is stronger than that 
of~\cite{Leane:2018kjk} (green dashed line), 
for $m_\chi \gtrsim 200\,{\rm GeV}$. 
The purple line is the unitarity upper bound on total 
$\langle \sigma v \rangle$~\cite{Griest:1989wd,Smirnov:2019ngs} 
and the black dashed line represents the cross-section required 
by a single-component WIMP to achieve $\Omega_\chi\,h^2 = 0.12$~\cite{Steigman:2012nb,Planck:2018vyg}. 
See the text for details.} 
\label{fig:vistot}
\end{figure}

% In Fig.~\ref{fig:vistot}, we show the maximum allowed total 
% $\langle \sigma v \rangle$ (at $95\%$ C.L.) obtained using 
% the data of all %previously mentioned 
% observations, but assuming the BR of $\nu\bar{\nu}$ 
% to be zero (green solid line). Here, CMB data play the 
% decisive role for $m_\chi \lesssim 5\,{\rm GeV}$, Fermi 
% and AMS-02 data are most stringent for 
% $5\,{\rm GeV} \lesssim m_\chi \lesssim 200\,{\rm GeV}$, 
% and for $m_\chi \gtrsim 200\,{\rm GeV}$ H.E.S.S 
% observation is most constraining. This limit lies 
% below the thermal relic cross-section (the black 
% dashed line~\cite{Steigman:2012nb}) up to 
% $m_\chi \sim 18\,{\rm GeV}$, signifying that for 
% $m_\chi \lesssim 18\,{\rm GeV}$, thermal WIMPs 
% annihilating into visible SM particle pairs only, 
% are ruled out by the existing data. The theoretical 
% upper bound on the total $\langle \sigma v \rangle$ 
% coming from the $S$-matrix unitarity is shown by the 
% purple line~\cite{Griest:1989wd,Smirnov:2019ngs}. 
% We have also shown the upper limit on the total 
% $\langle \sigma v \rangle$ obtained in~\cite{Leane:2018kjk} 
% (green dashed line). Note that, due to the 
% inclusion of the H.E.S.S data and other updated data 
% sets, our limit is stronger than that obtained 
% in~\cite{Leane:2018kjk} for $m_\chi \gtrsim 200\,{\rm GeV}$, 
% resulting in a smaller allowed region in the 
% $\langle \sigma v \rangle - m_\chi$ plane. 

Considering DM annihilations to visible 
SM channels, %\sout{we obtain} 
the $95\%$ C.L. upper limit on total 
$\langle \sigma v \rangle$, allowed by 
all existing data (green solid line in 
Fig.~\ref{fig:vistot}), is obtained. 
%\sout{the data of all observations.} 
Here, decisive roles are 
played by the CMB data for $m_\chi \lesssim 5\,{\rm GeV}$, 
Fermi-LAT and AMS-02 data for 
$5\,{\rm GeV} \lesssim m_\chi \lesssim 200\,{\rm GeV}$, 
and H.E.S.S data for $m_\chi \gtrsim 200\,{\rm GeV}$. 
The black dashed line represents the thermal relic 
cross-section for a single-component WIMP, 
which is larger than the maximum allowed 
$\langle \sigma v \rangle$ for 
$m_\chi \lesssim 18\,{\rm GeV}$. 
%\sout{This implies that} 
Thermal WIMPs annihilating into visible SM channels, 
are thus, ruled out up to $m_\chi \sim 18\,{\rm GeV}$. 
Unitarity upper bound on the total 
$\langle \sigma v \rangle$ is shown by the purple 
line~\cite{Griest:1989wd,Smirnov:2019ngs}. 
Comparison of our limit with that of~\cite{Leane:2018kjk} 
(green dashed line in Fig.~\ref{fig:vistot}) shows 
that, inclusion of H.E.S.S data improves the 
limits on total $\langle \sigma v \rangle$ 
for $m_\chi \gtrsim 200\,{\rm GeV}$, so that a smaller 
region in the $\langle \sigma v \rangle - m_\chi$ 
plane remains allowed.

\begin{figure*}[htb!]
\centering
\includegraphics[width=12.3cm,height=7.8cm]{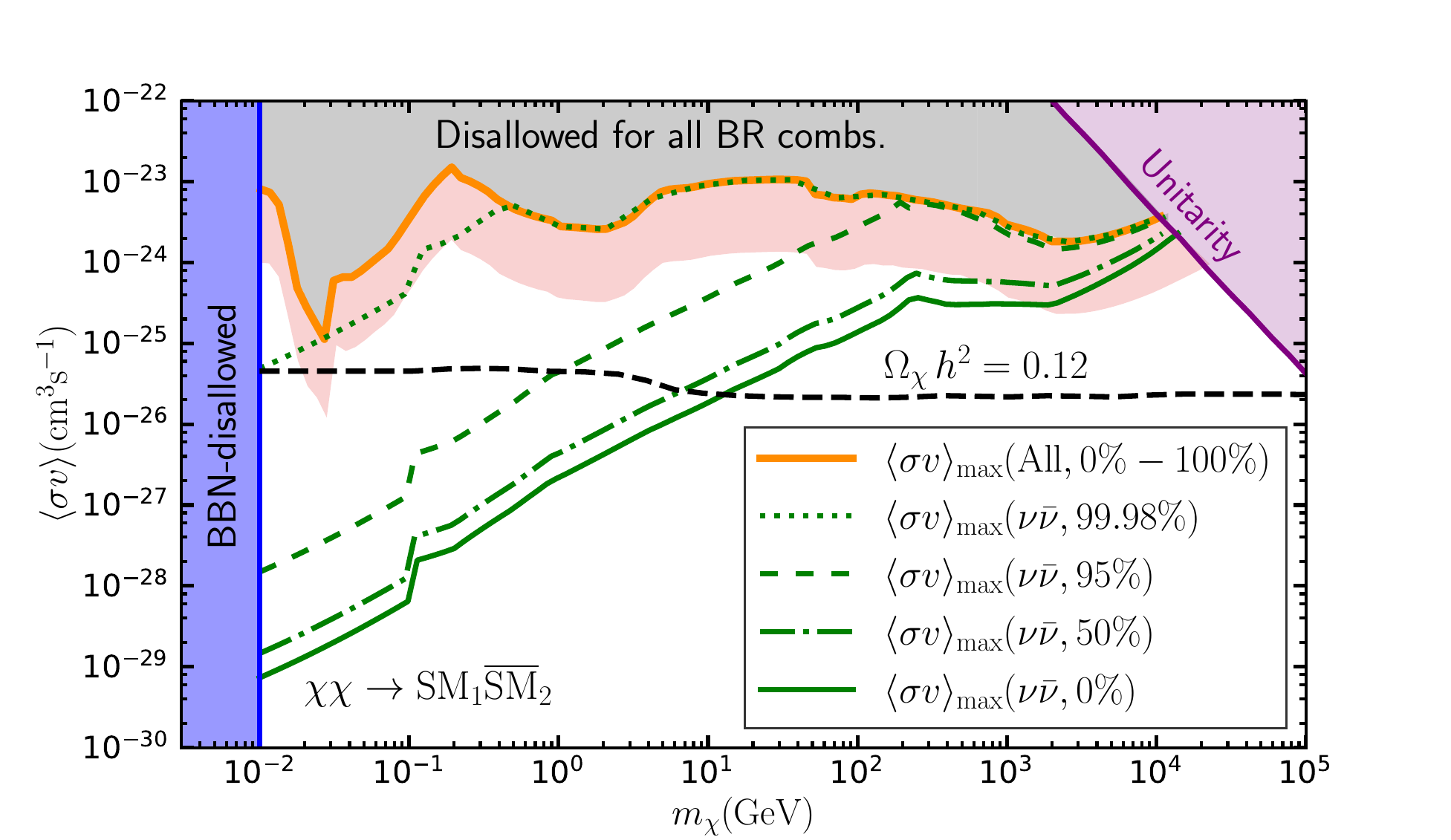}
\caption{For a single-component thermal WIMP 
in the $m_\chi$ range 10 MeV - 100 TeV, 
constituting all the observed DM, the orange line 
represents the BR-independent upper limit on total 
$\langle \sigma v \rangle$ (at $95\%$ C.L.) and 
the light red band shows its variation with the 
astrophysical uncertainties. 
The gray region is ruled out for all possible BR 
combinations, while, the blue region is 
disallowed by BBN~\cite{Nollett:2013pwa, Nollett:2014lwa}. 
The purple and the black dashed lines are the same as 
in Fig.~\ref{fig:vistot}. Variation of the maximum allowed 
total $\langle \sigma v \rangle$ with the BR 
attributed to $\nu\bar{\nu}$ are shown by 
the green lines. See the text for details. 
}
\label{fig:alltot}
\end{figure*}

% \sout{Next, we include the possibility of WIMP 
% annihilations into $\nu\bar{\nu}$.
% We scan over all possible BR combinations 
% (i.e., $0\% - 100\%$) of the kinematically 
% allowed SM channels and obtain the orange line shown 
% in Fig.~\ref{fig:alltot}, which represents the 
% maximum allowed total $\langle \sigma v \rangle$ for a 
% thermal WIMP, retaining the consistency with all existing data.} 
While considering WIMP annihilations into 
$\nu\bar{\nu}$, we scan over 
all possible BR combinations (i.e., $0\% - 100\%$) 
of the kinematically allowed SM channels and obtain 
the orange line shown in Fig.~\ref{fig:alltot}, 
which represents the maximum allowed total 
$\langle \sigma v \rangle$ for a thermal WIMP, 
retaining the consistency with all existing data. 
Note that, this limit is dictated by the Super-Kamiokande 
data of low energy neutrino search for $m_\chi \lesssim 100\,{\rm MeV}$, 
Super-Kamiokande, IceCube and ANTARES data of 
MW halo neutrino observation 
for $100\,{\rm MeV} \lesssim m_\chi \lesssim 800\,{\rm GeV}$ and 
H.E.S.S data of the Galactic halo $\gamma$-ray observation for 
$m_\chi \gtrsim 800\,{\rm GeV}$. 
For all $m_\chi$, this largest allowed 
$\langle \sigma v \rangle$ closely follows the upper 
limit obtained for exactly $100\%$ BR attributed 
to $\nu\bar{\nu}$. This maximum total $\langle \sigma v \rangle$ 
comes down for smaller values of $\nu\bar{\nu}$ BRs 
(the green lines), but still stays above 
the black dashed line for significant parts of the 
parameter space. The gray region is ruled out 
for all possible BR combinations, and the blue 
region is disallowed by BBN~\cite{Nollett:2013pwa,Nollett:2014lwa}. 

%\sout{One point is worth emphasizing here.} 
We emphasize that, in obtaining 
these maximum $\langle \sigma v \rangle$ 
(orange and green lines) the DM induced fluxes are calculated 
assuming a single-component WIMP constitutes all the DM 
in the galaxies. On the other hand, the black dashed curve 
signifies the minimum $\langle \sigma v \rangle$ 
allowed for a single-component DM that does not 
overclose the Universe. Therefore, the intermediate 
region below the orange curve and above the black 
dashed line is \textit{always allowed}. 
However, for a point lying within this region of the 
$\langle \sigma v \rangle - m_\chi$ plane, the relic 
density is in fact undersaturated by $\chi$. 
The resulting dilution of the annihilating DM flux, moves 
the maximum allowed $\langle \sigma v \rangle$ for any 
$m_\chi$, above the orange line. But this does not affect the 
general conclusion: 
%\textit{a general WIMP DM candidate 
%in the entire mass range 10 MeV - 100 TeV, which annihilates 
%into two-body SM final states, is allowed by all 
%currently available data.} 
%\textit{\sout{a general WIMP DM candidate 
%annihilating into two-body SM final states, with 
%mass in the range 10 MeV - 100 TeV, is allowed %by all currently available data.}}
\textit{a WIMP DM candidate 
annihilating into two-body SM final states 
(via $s$-wave processes), with mass 
in the range 10 MeV - 100 TeV, can be allowed 
by all currently available data.} 

Note that, this allowed mass range for a thermal WIMP changes with the BR attributed to $\nu\bar{\nu}$, as 
is evident from the green lines shown in Fig.~\ref{fig:alltot}. 
As we decrease the BR into $\nu\bar{\nu}$, kinematically 
allowed visible channels start contributing to the DM 
induced fluxes, thereby strengthening the upper limit 
on total $\langle \sigma v \rangle$. For example, if 
one assigns only $\sim 0.02\%$ BR to visible channels, 
the constraint on total $\langle \sigma v \rangle$ 
becomes stronger (dotted line), but the entire mass range 
10 MeV - 100 TeV still remains allowed. If the BR 
of $\nu\bar{\nu}$ is decreased further, the upper 
limit on total $\langle \sigma v \rangle$ gradually 
strengthens, until all $m_\chi$ lower than $\sim 18\,{\rm GeV}$ 
become ruled out for DM annihilations to only visible 
SM channels (solid line). 
Therefore, it is evident that WIMPs of 
$m_\chi \lesssim \mathcal{O}(10\,{\rm GeV})$ 
are allowed only if they dominantly annihilate 
into SM neutrinos. Such a feature can be 
realized by invoking extra symmetries and light 
sterile neutrinos which significantly couple to 
both the SM neutrinos and the 
DM~\cite{Blennow:2019fhy,Ballett:2019pyw,Dutta:2019fxn,ElAisati:2017ppn,Farzan:2011ck}.  

It is to be noted that the dependency 
of our results on the values of the astrophysical 
parameters is also explored. Since the choice 
of $D(\rho,|\vec{r}|,z)$ does not affect the 
limits considerably (see Fig.~\ref{fig:AMS02ep}; 
right panel), we keep it fixed at 
$D_0 = 2.7\times 10^{28}\,{\rm cm}^2{\rm s}^{-1}$, 
$\delta = 0.6$ and $z_t = 4\,{\rm kpc}$, 
%\sout{the value mentioned earlier,} 
but vary $B_\odot$ in the range 
$5.7 - 8.9\,\mu{\rm G}$~\cite{John:2021ugy,Leane:2018kjk}, 
$\Phi$ within $0.46 - 0.6\,{\rm GV}$~\cite{Cholis:2015gna} 
and $\rho_\odot$ in the range $0.25 - 0.7\,{\rm GeV}\,{\rm 
cm}^{-3}$~\cite{Salucci:2010qr,Bovy:2012tw,Benito:2019ngh,Benito:2020lgu}. 
On the other hand, we have kept the values 
of the dSph J-factors fixed at the values given 
in~\cite{Fermi-LAT:2016uux}. 
The resulting BR-independent upper limits always 
lie within the light red band 
(see Fig.~\ref{fig:alltot}) which is above the 
black dashed line for most of the $m_\chi$ 
values in the $10\,{\rm MeV} -100\,{\rm TeV}$ range. 
We find that the variation of $\rho_\odot$ has the 
most prominent impact on this maximum allowed 
total $\langle \sigma v \rangle$.

%\vspace{1mm}
%\noindent
\section{Conclusions}
\label{sec:sec5}

In this work, we have obtained the most 
conservative BR-independent upper limit on
the total $\langle \sigma v \rangle$ of a 
10 MeV - 100 TeV thermal WIMP, annihilating into 
all possible SM final states via $2 \rightarrow 2$
$s-$wave processes, using the most updated data from several
astrophysical and cosmological observations, i.e, 
Planck~\cite{Planck:2015fie,Planck:2018vyg}, 
Fermi-LAT~\cite{Fermi-LAT:2016uux}, AMS-02~\cite{AMS:2019rhg}, 
H.E.S.S~\cite{HESS:2022ygk}, 
Super-Kamiokande~\cite{Super-Kamiokande:2020sgt,Super-Kamiokande:2021jaq}, 
IceCube~\cite{IceCube:2017rdn} and ANTARES~\cite{Albert:2016emp}. 
In deriving this constraint, we vary the BRs of 
all kinematically allowed annihilation channels arbitrarily 
in the range $0\%-100\%$ and find that the maximum allowed total 
$\langle \sigma v \rangle$ lies in the range 
$10^{-23}\,- 10^{-25}\,{\rm cm}^{3}\,{\rm s}^{-1}$ for 
the entire $m_\chi$ range considered. This limit 
%is sensitive to the values of the astrophysical parameters and 
strengthens at most by an order of magnitude with the 
variations of the astrophysical parameters. We note that, 
this upper limit on the total $\langle \sigma v \rangle$ 
almost always coincides with the upper limit obtained for 
exactly $100\%$ BR attributed to $\nu\bar{\nu}$. As the 
BR of $\nu\bar{\nu}$ decreases, the constraint gradually 
strengthens and for annihilations into visible final states 
only, thermal WIMPs are ruled out up to $\sim 18$ GeV. 
%for annihilations into visible SM particles. 
Therefore, for $m_\chi \lesssim \mathcal{O}(10\,{\rm GeV})$, 
WIMPs are allowed provided they dominantly annihilate into 
SM neutrinos. 
%\red{\bf Particle physics models endowed with such 
%feature are available in the literature 
%(see~\cite{Blennow:2019fhy,Ballett:2019pyw,Dutta:2019fxn,ElAisati:2017ppn,Farzan:2011ck}), 
%where our results are readily applicable.}
In addition, the importance of H.E.S.S data in 
constraining multi-TeV WIMPs is also emphasized. It is 
the inclusion of H.E.S.S data which strengthens our limit 
compared to that obtained in~\cite{Leane:2018kjk}, for 
$m_\chi \gtrsim 200\,{\rm GeV}$. Several future generation 
observations~\cite{https://doi.org/10.48550/arxiv.1306.6175,Cumani:2015ava,Engel:2022bgx,Pierre:2014tra,Silverwood:2014yza,DAMPE:2017cev,DiSciascio:2016rgi,PhysRevD.99.021302,Kar:2019cqo,IceCube:2019pna,Baur:2019jwm,Lopez-Coto:2022pff,KM3Net:2016zxf,Fermani:2020oxx,Lopez-Coto:2022xsm,Hyper-Kamiokande:2018ofw,Bell:2020rkw,Robles:2022hzj,DUNE:2018tke,DUNE:2018hrq,DUNE:2018mlo,Miranda:2022kzs,Akita:2022lit,Arguelles:2019xgp}, 
in particular, the neutrino 
observations~\cite{IceCube:2019pna,Baur:2019jwm,Lopez-Coto:2022pff,KM3Net:2016zxf,Fermani:2020oxx,Lopez-Coto:2022xsm,Hyper-Kamiokande:2018ofw,Bell:2020rkw,Robles:2022hzj,DUNE:2018tke,DUNE:2018hrq,DUNE:2018mlo,Miranda:2022kzs,Akita:2022lit,Arguelles:2019xgp} and
the observations that can constrain 
$\rho_\odot$~\cite{Gaia:2016lbo,Gaia:2018ydn}, will improve 
the constraints on the total $\langle \sigma v \rangle$ 
of thermal WIMPs.

Note that, instead of two-body annihilation channels if 
the WIMP possesses multi-body annihilation channels, 
including those involving dark sector 
particles~\cite{Pospelov:2007mp,Berlin:2016gtr,Evans:2017kti,Hooper:2019xss}, 
the upper-limits on the total $\langle \sigma v \rangle$ 
%annihilation cross-section 
are even weaker~\cite{Ibarra:2012dw,Elor:2015tva,Elor:2015bho}.
Furthermore, for WIMP annihilations dominated by $p$-wave 
processes, $\langle \sigma v \rangle$ is proportional to the 
square of the DM relative velocity, whose value in the present 
day galaxies is suppressed by a few orders of magnitude compared 
to its value at freeze-out. Hence, in such cases, the relic 
density constraint becomes less restrictive, in terms of the 
present value of $\langle \sigma v \rangle$~\cite{PhysRevD.99.061302,Arguelles:2019ouk}~\footnote{Although the 
annihilation cross-section of fermionic scalar portal DM is 
p-wave suppressed, their scattering cross-section with protons 
is unsuppressed~\cite{Lopez-Honorez:2012tov}, thereby leading 
to direct search constraints on the parameter space 
of such DM candidate~\cite{Arcadi:2019lka,Arcadi:2021mag}.}.
A similar relaxation is expected if co-annihilations 
contribute to thermal freeze-out. These further 
strengthen our claim that, in general, thermal WIMPs 
in the entire 10 MeV - 100 TeV range, are still 
consistent with all observational data.

\vspace{1cm}
%\noindent
%\begin{Acknowledgments}
\section*{Acknowledgements}
The work of K.D. is partially supported by the Indo-Russian grant 
DST/INT/RUS/RSF/P-21, MTR/2019/000395, and Core Research Grant 
CRG/2020/004347 funded by SERB, DST, Government of India. 
The research of A.K. was supported by the National
Research Foundation of Korea(NRF) funded by the Ministry of Education
through the Center for Quantum Space Time (CQUeST) with grant number
2020R1A6A1A03047877 and by the Ministry of Science and ICT with grant
number 2021R1F1A1057119.	
%\end{Acknowledgments}

%************************************************************
%\bibliographystyle{apsrev4-1}
%\bibliography{refs}

\providecommand{\href}[2]{#2}
\addcontentsline{toc}{section}{References}
\bibliographystyle{JHEP}
	
\bibliography{refs}

\end{document}